# Development of a novel nonlinear dynamic cavitation model and its numerical validations

Haidong Yu[12], Xiaobo Quan[3], Haipeng Wei[1], Matevž Dular[4] and Song Fu[2*]
[1]Beijing Institute of Astronautical System Engineering, Beijing, 100076, China
[2]School of Aerospace Engineering, Tsinghua University, Beijing 100084, China
[3]China Academy of Launch Vehicle Technology, Beijing, 100076, China
[4]Laboratory for Water and Turbine Machines, University of Ljubljana, Aškerčeva 6, 1000 Ljubljana, Slovenia

## Abstract

Aiming at modeling the cavitation bubble cluster, we propose a novel nonlinear dynamic cavitation model (NDCM) considering the second derivative term in Rayleigh-Plesset equation through strict mathematical derivation. There are two improvements of the new model: i) the empirical coefficients are eliminated by introduction of the nonuniform potential functions of $\psi_v$ and $\psi_c$ for growth and collapse processes respectively, and ii) only two model parameters are required, which both base on physical quantities – the Blake critical radius $R_b$ and the average maximum growth radius $R_m$. The corresponding cavitation solver was developed by using OpenFOAM in which we implemented the modified momentum interpolation (MMI) method to ensure that the calculated results are independent of time step size. Three validation cases, namely numerical bubble cluster collapse, ultrasonic horn experiment, and hydrodynamic cavitation around slender body are employed. The results indicate that $\psi_v$ and $\psi_c$ can reveal the nonlinear characteristics for cavity accurately, and $R_b$ and $R_m$ can reflect the relevance between cavitation model and actual physical quantities. Moreover, it is discussed the potentiality of NDCM that is generally applied on the cavitating flow possessing with dispersed bubbly cloud.

* Corresponding author, email: fs-dem@tsinghua.edu.cn

## Highlights

- A nonlinear dynamic cavitation model (NDCM) is established through strict mathematical derivation.
- The formula of NDCM employs the potential functions of $\psi_v$ and $\psi_c$ instead on empirical coefficients to describe the nonlinear effects.
- The model parameters of $R_b$ and $R_m$ represent the physical characteristics of bubble cluster.
- The NDCM is best valid for the cavitating flows with dispersed bubbles.

# 1. Introduction

Cavitation often occurs in liquid flows when the ambient pressure drops below a certain threshold. The cavitating bubbles will emerge gradually from “cracked” liquid medium at its weak points [1]. Individual bubbles cluster and form a complex two-phase mixture cloud, which shape depends strongly on the structure of the flow field. The cavitation bubble cluster exhibits many unique characteristics, such as strong collapse accompanied by a shockwave, or the natural frequency far lower than single bubble’s etc. [2, 3] There are many researches on single bubble dynamics which can be described by Rayleigh-Plesset type equation [4, 5]. However, an approach to build cavitation model based on Rayleigh-Plesset equation to investigate the dynamics of bubble cluster can be considered questionable.

Numerical simulation of cavitating flows and specifically the development of transport equation model (TEM) has received enormous attention from investigators in recently years. Instead of potential flow theory implemented in early engineering applications, the Eulerian's one field formulation (OFM) of the two-phase Navier–Stokes equation [6-9], which combines the properties of each phase as a single mixed one, is popularly applied as the methodology of multiphase model. The cavitation model for bubble cluster is embedded into the convective phase equation as source terms. There is also an alternative approach based on Eulerian–Lagrangian method [10, 11] but is not within the present framework.

The prototype of TEM was introduced by Kubota [12] who assumed that the bubble nuclei are uniformly distributed in the flow, and the simplified Rayleigh-Plesset equation, which considered the SGS bubble interaction was used to determine the change of radius and consequently the mixture density in each computational cell. The advantage of this approach is that the dynamic response of local equilibrium bubbles can be estimated precisely. However, solving the nonlinear transport equation often faces the difficulty of convergence and its application was merely limited in studying steady flow.

In order to develop the Kubota’s method, the modeling strategy was focused on larger scale of bubble cloud rather than the tiny scale of individual bubbles. A different approach with the same form of transport equation models had already been proposed in literature [6-9], in which the nonlinear ODE is replaced by a convective equation of void fraction. The mass transfer rate of bubbles is contained implicitly into source terms. It has the advantage that the convective character of equation is more appropriate for describing the topological evolution of bubble cluster particularly in unsteady situations. In the literature, there are two main approaches to build the empirical phase-transition laws associating with surrounding pressure. Merkel et al. [13-15] proposed simplest formulation based on dimensional analysis that defined the characteristic velocity $u_\infty$ or temporal scale $t_\infty$. However, these parameters are prescribed as a constant during the whole dynamic period, especially in collapsing process, which would cause large deviation. In order to make the model more physical, the void fraction was regarded equivalently as a group of identical bubbles. Schnerr et al. [16-19] simplified the Rayleigh-Plesset equation that only the linear part was used to redefine the velocity scale as $\sqrt{2|p-p_{sat}|/3\rho_l}$ for both growth and collapse situations. The advantage of this formulation is that it follows, to some part, a physical law and reduces the complexity. Nevertheless, the empirical coefficients are still needed to regulate the order of magnitude between evaporation and condensation terms.

Another way of modelling is to establish the barotropic relation that couples the mixture density and local pressure as called equation of state (EoS). The model was first introduced by Delannoy and Kueny [20] and later widely used by others [21-23]. Some results obtained by EoS model show good agreement with benchmarks in term of strong compressible multiphase flows. However, the mechanisms inside cavity are complicated more than barotropic assumption that the gradient parallel between pressure and density can hardly be guaranteed. For weak compressible issues, numerical algorithms lost robustness that would induce numerical instability and sometimes poor convergence.

All the mentioned cavitation models were developed and compared with a variety of experiments. Despite some satisfactory results have been achieved by existing models, some questions still remain to be discussed. One of the basic principles for cavitation model application is that the flow pattern of vapor-liquid mixture should be homogeneous dispersed bubbly flow. It will exceed the model capability if the frequency of bubble coalescence is too high, for instance, in the case of supercavitating flow. Though few models are still workable on the flow out of their application scope, one has to select weird values for parameters in order to match the results. Furthermore, it can be inferred from several simulation comparisons [24, 25] that the model parameters cannot reflect the physical characteristics sufficiently because the empirical coefficients in specified cavitation model sometimes need to be recalibrated for different flow conditions. Also, the results calculated by different cavitation models present respective tendencies under the same experimental condition [26, 27]. We speculate the main reasons causing these deviations come from two sides. One is that the underlying characteristic scales produce large errors when the second derivative term in Rayleigh-Plesset equation becomes predominant. As previously mentioned, the characteristic velocity in current models is given as an immutable value $u_\infty$ or linear form $\sqrt{2|p-p_{sat}|/3\rho_l}$ so that it is necessary to introduce empirical coefficients to compensate the information mismatch. On the other hand, the typical configuration of existing models with two empirical coefficients and several parameters have good adaptability but wide range of proper values to guess. Defining the parameters which can reveal the physical relevance between model and objective phenomenon would narrow the search regions. Thus, we suggest three improvements to account for the effect of nonlinear term, eliminating the empirical coefficients, and parameters reflected more intrinsic law to remedy the cavitation model with wider applicability.

In this paper, a novel TEM-kind nonlinear dynamic cavitation model is proposed with rigorous mathematical derivation. There arise two functions $\psi_v$ and $\psi_c$ instead of empirical coefficients to represent the nonlinear effects for growth and collapse periods respectively. Only two meaningful parameters, the Blake critical radius $R_b$ and the average maximum growth radius $R_m$, are adopted. The new model is implemented in the open source C++ package OpenFOAM based on the interPhaseChangeFoam solver [28]. To illustrate the superiorities of the new model clearly, three typical validations cases, sequentially from simple to complex, were adopted: numerical bubble cluster collapse [29], ultrasonic horn experiment [30, 31], and hydrodynamic cavitation around a slender body [32]. In these cases, the similar dispersed structures of bubble cloud satisfy the homogeneous assumption.

In the following sections, basic assumptions and simplifications for bubble cluster are stated firstly. The model derivation processes are then elaborated in detail. The algorithms, including VoF-based interface capture method MULES [33, 34] in OpenFOAM and modified momentum interpolation (MMI) method [35, 36] for multiphase flows, are implemented correspondingly. Finally, qualitative and quantitative validations of the improved model performance are presented.

# 2. Modeling methodology

## 2.1 Assumptions

The cavitating cluster composed by multi-radius bubbles makes the overall problem difficult to model. In order to reduce the complexity of modeling the bubble cloud, we propose to introduce four assumptions to describe its main mechanisms approximately.

I. No bubbles' coalescence or breakage.

II. Local homogeneous assumption:

As is shown in Fig. 1, the bubble cloud is divided into lattices by CFD grid in physical space. The real bubbles included in each grid cell are equivalent as uniform radius according to the local vapor volume so that the free degrees are degenerated into one.

III. Neglecting the local bubble interactions:

The intensity of bubble interactions has positive correlations with the bubble population $N$ [37]. Noting that only partial region of bubble cloud is filled in local grid cell. If the mesh resolution is fine enough to contain few bubbles in single element that the interactions between these localized bubbles could be ignored.

IV. The law of mean bubble dynamics:

Given a pressure field, bubbles with different scales experience similar dynamics which can be governed by Rayleigh-Plesset-type equations. We suppose the mean bubble dynamic processes existed which could be obtained from statistical operations on the whole cloud bubbles.

In local point of view, the assumptions II and III are indicated that the dynamics of average bubbles, including growth and collapse, insides each grid cell can be treated as synchronized processes. It should be noted that the assumption III doesn't mean nothing interaction existed inside bubble cloud. Actually, these interactions are implied in numerical fluxes between adjacent grid cells, which will be illustrated in section 4.1. It can also be inferred that there is a lower limit of mesh resolution that fulfills the requirements in assumption III. We recommend to generate a proper mesh by using grid independence principle.

In overall view of bubble cloud, the structure of void fraction which exhibits concentrated interior and sparse border has been observed through several experiments [38, 39]. This characteristic can be reflected as radius distribution of the average bubbles mentioned above in computational space. The assumption IV is indicated that such distribution can be decided by the mean bubble dynamics which is considered as internal law that every bubble obeys to manifest the main dynamic character of bubble cloud. Therefore, the modeling strategy can be simplified as to investigate the mean dynamics of local average bubbles.

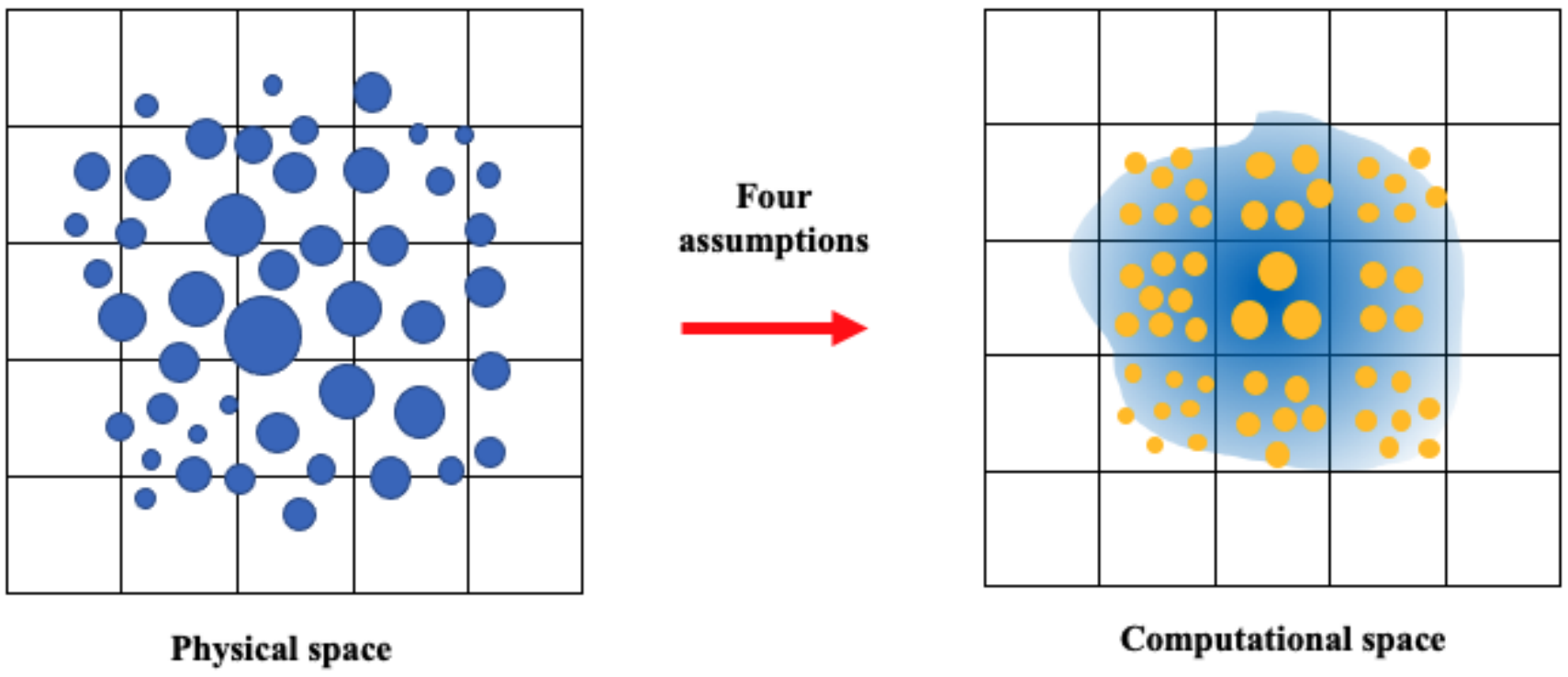

Fig. 1 Diagram of bubble cluster in physical/computational space

## 2.2 Simplifications

With the above assumptions, the model expressions can be derived from the individual bubble with Rayleigh–Plesset equation:

$$\underbrace{\frac{p_{sat}-\bar{p}}{\rho_l}}_{1}+\underbrace{\frac{p_{g0}}{\rho_l}\left(\frac{R_0}{R}\right)^{3\gamma}}_{2}=R\underbrace{\frac{d^2R}{dt^2}}_{3}+\underbrace{\frac{3}{2}\left(\frac{dR}{dt}\right)^2}_{4}+\underbrace{\frac{2\sigma}{\rho_l R}}_{5}-\underbrace{\frac{4\mu}{\rho_l R}\frac{dR}{dt}}_{6}. \tag{1}$$

Here the first term represents the difference of vapor saturation pressure $p_{sat}$ and ambient pressure $\bar{p}$. In Euler mixture model, we should point out that $\bar{p}$ is a phase-weighted mixing pressure [6] which can be considered as ambient pressure for those under-resolved bubbles. The second terms describe the influence of gases inside the bubble. The inertial effects are given by the third and the fourth terms, while terms 5 and 6 depict the influence of surface tension and liquid viscosity, respectively.

Schnerr [17] has introduced the bubble density $n$ which definition represents the bubble population per unit liquid volume is adopted uniformly to build the relation between liquid volume fraction $\alpha_l$ and bubble radius:

$$\alpha_l = \frac{V_l}{V_l + \frac{4}{3}\pi R^3 n V_l} = \frac{1}{1 + \frac{4}{3}\pi R^3 n}. \tag{2}$$

Due to the cavitating bubble is commonly treated as pure vapor, the expression for mass transfer rate per unit volume can be built as follows:

$$\dot{m} = \frac{1}{V}\frac{d}{dt}\left(\rho_v * \frac{4}{3}\pi R^3 * n\alpha_l V\right) = \rho_v \alpha_l^2 \frac{d}{dt}\left(\frac{4}{3}\pi n R^3\right). \tag{3}$$

According to the meaning of $n$, the non-dimensional term $4\pi n R^3/3$ inside time derivative indicates the volume ratio of vapor to liquid. The original way to handle this term is to expand differentially into $4\pi R^2\, dR/dt$, then specify the velocity $dR/dt$ by using the linear part $\sqrt{2|p - p_{sat}|/3\rho_l}$ in Rayleigh-Plesset equation for both growth and collapse processes. Except for the oversimplification of linearized bubble velocity as previously discussed, such instantaneous form would underestimate the intensity of source terms during time marching. Thus, we suggest to utilize the average value within the evolution time $\tau$ (Eqn.4) during which the bubble radius would vary from present $R_P$ to ultimate $R_U$:

$$\frac{d}{dt}\left(\frac{4}{3}\pi n R^3\right) \approx n * \frac{\frac{4}{3}\pi R_U^3 - \frac{4}{3}\pi R_P^3}{\tau}. \tag{4}$$

It is illustrated the implication of $\tau$ through the numerical results of Rayleigh-Plesset equation. As is shown in Fig. 2, neglecting the bubble rebound, the typical dynamic period can be divided into four stages, namely inception, growth, slow-down and collapse, also several critical moments are marked - including the Blake radius $R_b$, zero acceleration point $R_d$, maximum radius $R_m$, and the final collapse radius $R_{c0}$. The dynamics before $R_b$ is gas dominated oscillation, after that the bubble content is occupied by vapor gradually, then unstable cavitation took place. The expansion process from $R_b$ to $R_m$ is split by inflection point $R_d$ into growth and slow-down, where the ratio of $R_d$ and $R_m$ is approximately 0.8. Besides, we emphasize that the nonlinear effect should be considered for the accelerated collapse. The cavitation model is only devoted to describing growth and collapse so the evolution time for these two periods, $\tau_v$ and $\tau_c$, are determined from the present point $R$ to $R_d$ and $R_{c0}$, respectively.

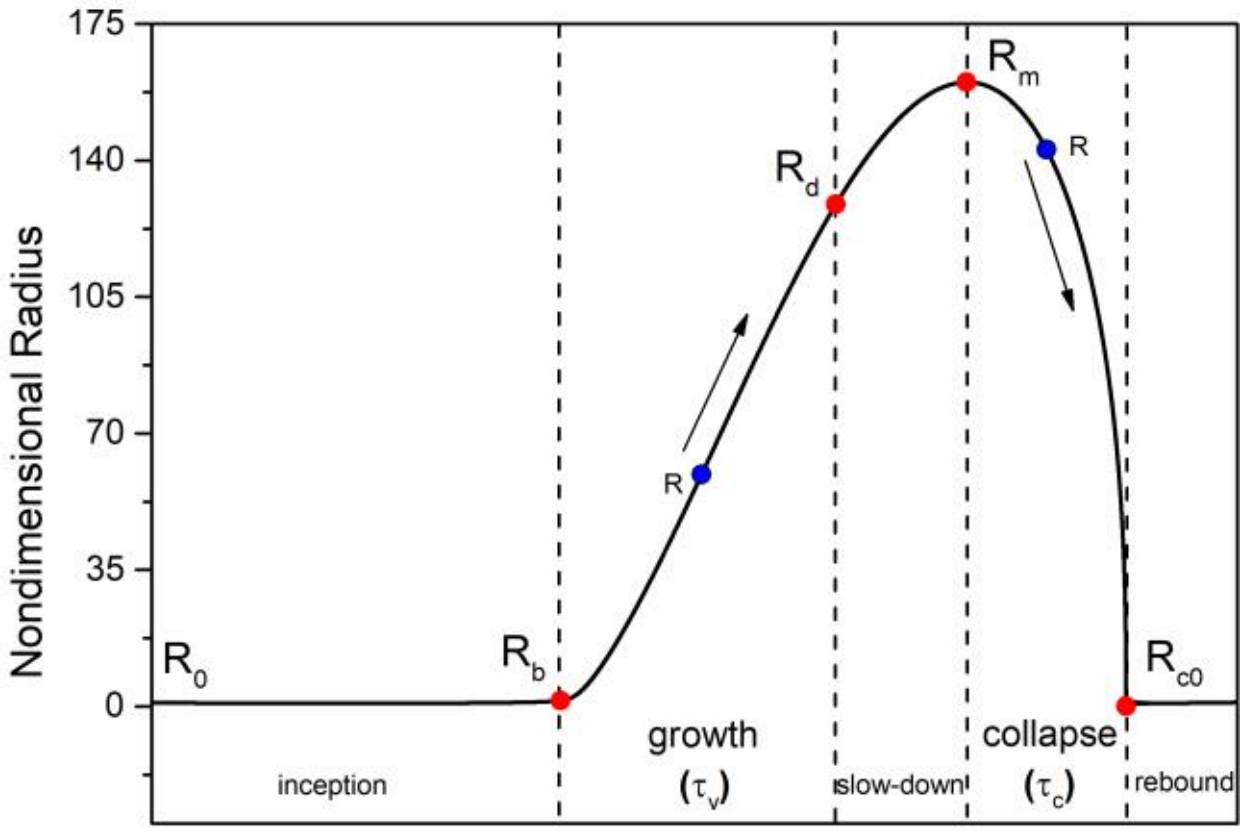

Fig. 2 Typical bubble dynamic period

Therefore, the main task is to calculate the values of $\tau_v$ and $\tau_c$ in term of Rayleigh–Plesset equation (Eqn.1) where terms including surface tension, viscosity and gas had been neglected,

$$R\ddot{R} + \frac{3}{2}\dot{R}^2 = \frac{p_{sat} - \bar{p}}{\rho_l}. \tag{5}$$

Here, the saturated vapor pressure $p_{sat}$ and liquid density $\rho_l$ are treated as constants, and the LHS in Eqn.5 can also be reformulated as the following equation.

$$R\ddot{R} + \frac{3}{2}\dot{R}^2 = \frac{1}{2R^2\dot{R}}\frac{d}{dt}\left(\dot{R}^2R^3\right). \tag{6}$$

It can be expected that the exact solutions for these evolution time can be derived from integrating the bubble velocity $dR/dt$, i.e., to integrate the Eqn.5 twice under compatible initial conditions. The deductions are discussed in next section.

## 2.3 Modeling

### 2.3.1 Growth potential function $\dot{m}_v$

Based on the interval of growth period shown in Fig. 2, the integration and corresponding initial condition for obtaining the velocity at $R$ are given in Eqn.7.1 and 7.2, where the velocity at $R_b$ is treated approximately zero:

$$t = 0, R = R_b \; and \; \dot{R} \approx 0, \tag{7.1}$$

$$\int_0^{t_R} \frac{d}{dt}\left(\dot{R}^2R^3\right) dt = \frac{p_{sat} - \bar{p}}{\rho_l} * \int_0^{t_R} 2R^2\frac{dR}{dt}\, dt. \tag{7.2}$$

Assuming that the ambient pressure $\bar{p}$ is constant during short period of time, the integral result of growing velocity is then given in Eqn.8:

$$\frac{dR}{dt} = \sqrt{\frac{2}{3}\frac{p_{sat} - \bar{p}}{\rho_l}\left(1 - \left(\frac{R_b}{R}\right)^3\right)}. \tag{8}$$

Obviously, it approaches to the inertial velocity $\sqrt{2|\bar{p} - p_{sat}|/3\rho_l}$ when the bubble expands significantly greater than $R_b$. Consequently, the evolution time of $\tau_v$ can be integrated from $R$ to $R_d$ using Eqn.8:

$$\tau_v = \int_R^{R_d} \frac{dR}{\sqrt{\frac{2}{3}\frac{p_{sat}-\bar{p}}{\rho_l}\left(1-\left(\frac{R_b}{R}\right)^3\right)}} = \frac{1}{\sqrt{\frac{2}{3}\frac{p_{sat}-\bar{p}}{\rho_l}}}\int_R^{R_d} \frac{dR}{\sqrt{1-\left(\frac{R_b}{R}\right)^3}}. \tag{9}$$

Letting $x = (R_b/R)^3$, so we have:

$$dR = -\frac{R_b}{3}x^{-\frac{4}{3}}dx. \tag{10}$$

Thus, Eqn.9 can be reformulated as follows:

$$\tau_v = \frac{R_b}{3\sqrt{\frac{2}{3}\frac{p_{sat}-\bar{p}}{\rho_l}}}\int_{\left(\frac{R_b}{R_d}\right)^3}^{\left(\frac{R_b}{R}\right)^3} x^{-\frac{4}{3}}(1-x)^{-\frac{1}{2}}dx, \tag{11}$$

where the definite integral in Eqn.11 can be expanded by applying the Gauss hypergeometric function, the result becomes:

$$\tau_v = \frac{2R_b}{3\sqrt{\frac{2}{3}\frac{p_{sat}-\bar{p}}{\rho_l}}}\left[\begin{array}{l}\sqrt{1-\left(\frac{R_b}{R_d}\right)^3}\ {}_2F_1\left(\frac{1}{2},\frac{4}{3};\frac{3}{2};1-\left(\frac{R_b}{R_d}\right)^3\right)\\ -\sqrt{1-\left(\frac{R_b}{R}\right)^3}\ {}_2F_1\left(\frac{1}{2},\frac{4}{3};\frac{3}{2};1-\left(\frac{R_b}{R}\right)^3\right)\end{array}\right]. \tag{12}$$

Note that there is an ultimate limit under the large ratio of $R_d$ an $R_b$, then:

$$\lim_{\frac{R_b}{R_d}\to 0}\left(\frac{R_b}{R_d}\right)\sqrt{1-\left(\frac{R_b}{R_d}\right)^3}\ {}_2F_1\left(\frac{1}{2},\frac{4}{3};\frac{3}{2};1-\left(\frac{R_b}{R_d}\right)^3\right) = \frac{3}{2}. \tag{13}$$

The Eqn.12 is simplified as follow:

$$\tau_v = \frac{2}{3\sqrt{\frac{2}{3}\frac{p_{sat}-\bar{p}}{\rho_l}}}\left[\frac{3}{2}R_d - R_b\sqrt{1-\left(\frac{R_b}{R}\right)^3}\ {}_2F_1\left(\frac{1}{2},\frac{4}{3};\frac{3}{2};1-\left(\frac{R_b}{R}\right)^3\right)\right]. \tag{14}$$

Thus, combing with the Eqn.2 and Eqn.14, we can get the approximation of derivative in Eqn.4 for growth situation:

$$\frac{d}{dt}\left(\frac{4}{3}\pi n R^3\right) \approx n * \frac{\frac{4}{3}\pi R_d^3 - \frac{4}{3}\pi R^3}{\tau_v} = \frac{1-\alpha_l}{\alpha_l}\frac{1}{R}\sqrt{\frac{2}{3}\frac{p_{sat}-\bar{p}}{\rho_l}} * \psi_v, \tag{15.1}$$

$$\psi_v = \frac{\left(\frac{R_d}{R}\right)^3 - 1}{\left(\frac{R_d}{R}\right) - \frac{2}{3}\left(\frac{R_b}{R}\right)\sqrt{1-\left(\frac{R_b}{R}\right)^3}\ {}_2F_1\left(\frac{1}{2},\frac{4}{3};\frac{3}{2};1-\left(\frac{R_b}{R}\right)^3\right)}. \tag{15.2}$$

There, it emerges the formulation $\psi_v$ which we call growth potential function that represents the capacity of continuous growth until $R_d$ being reached. Recalling the ratio of $R_d/R_m \approx \eta = 0.8$, it can be expressed by $R_m$ alternatively:

$$\psi_v \cong \frac{\left(\eta\frac{R_m}{R}\right)^3 - 1}{\left(\eta\frac{R_m}{R}\right) - \frac{2}{3}\left(\frac{R_b}{R}\right)\sqrt{1-\left(\frac{R_b}{R}\right)^3}\ {}_2F_1\left(\frac{1}{2},\frac{4}{3};\frac{3}{2};1-\left(\frac{R_b}{R}\right)^3\right)}. \tag{16}$$

2.3.2 Collapse potential function $\dot{m}_c$

Analogously, with the same integration in Eqn.7.2, the initial condition at max radius $R_m$ for obtaining the collapse velocity is given as follows:

$$t = 0, R = R_m \ and\ \dot{R} = 0, \tag{17}$$

which yields

$$\frac{dR}{dt} = -\sqrt{\frac{2}{3}\frac{\bar{p}-p_{sat}}{\rho_l}\left(\left(\frac{R_m}{R}\right)^3 - 1\right)}. \tag{18}$$

The evolution time $\tau_c$ can be integrated from $R$ to $R_{c0}$ by Eqn.18:

$$\tau_c = -\int_R^{R_{c0}} \frac{dR}{\sqrt{\frac{2}{3}\frac{\bar{p}-p_{sat}}{\rho_l}\left(\left(\frac{R_m}{R}\right)^3 - 1\right)}} = \frac{1}{\sqrt{\frac{2}{3}\frac{\bar{p}-p_{sat}}{\rho_l}}}\int_{R_{c0}}^{R} \frac{dR}{\sqrt{\left(\frac{R_m}{R}\right)^3 - 1}}. \tag{19}$$

Letting $x = (R/R_m)^3$, so we have:

$$dR = \frac{R_m}{3}x^{-\frac{2}{3}}dx. \tag{20}$$

Eqn.19 can be reformulated as follows:

$$\tau_c = \frac{R_m}{3}\frac{1}{\sqrt{\frac{2}{3}\frac{\bar{p}-p_{sat}}{\rho_l}}}\int_{\left(\frac{R_{c0}}{R_m}\right)^3}^{\left(\frac{R}{R_m}\right)^3} x^{-\frac{1}{6}}(1-x)^{-\frac{1}{2}}dx. \tag{21}$$

According to the expression of incomplete beta function, we obtain:

$$\tau_c = \frac{R_m}{3}\frac{1}{\sqrt{\frac{2}{3}\frac{\bar{p}-p_{sat}}{\rho_l}}}\left[\beta\left(\frac{5}{6},\frac{1}{2},\left(\frac{R}{R_m}\right)^3\right) - \beta\left(\frac{5}{6},\frac{1}{2},\left(\frac{R_{c0}}{R_m}\right)^3\right)\right]. \tag{22}$$

Employing Eqn.2 and Eqn.22, we can get the approximation of derivative in Eqn.4 for collapse situation:

$$\frac{d}{dt}\left(\frac{4}{3}\pi n R^3\right) \approx n * \frac{\frac{4}{3}\pi R_{c0}^3 - \frac{4}{3}\pi R^3}{\tau_c} = -\frac{1-\alpha_l}{\alpha_l}\frac{1}{R}\sqrt{\frac{2}{3}\frac{\bar{p}-p_{sat}}{\rho_l}} * \psi_c, \tag{23.1}$$

$$\psi_c = \frac{3\left(\frac{R}{R_m}\right)\left[1-\left(\frac{R_{c0}}{R}\right)^3\right]}{\beta\left(\frac{5}{6},\frac{1}{2},\left(\frac{R}{R_m}\right)^3\right) - \beta\left(\frac{5}{6},\frac{1}{2},\left(\frac{R_{c0}}{R_m}\right)^3\right)}. \tag{23.2}$$

Emerged $\psi_c$ is called collapse potential function considering the historical effect that reflects the magnitude of bubble velocity applied by continuously accelerated.

It is further proposed that the function $\psi_c$ can be simplified by constraining the bubble radius $R$ above the initial radius, where the last moment at $R_{c0}$ is actually at least one magnitude less than the initial state. This yield:

$$\psi_c \cong \frac{3\left(\frac{R}{R_m}\right)}{\beta\left(\frac{5}{6},\frac{1}{2},\left(\frac{R}{R_m}\right)^3\right)}. \tag{24}$$

Based on analysis above, the final form of NDCM is given below by replacing Eqn.15.1 and Eqn.23.1 into Eqn.3:

$$\dot{m}_v = \rho_v\alpha_l(1-\alpha_l)\frac{1}{R}\psi_v\sqrt{\frac{2}{3}\frac{p_{sat}-\bar{p}}{\rho_l}} \quad (p_{sat} > \bar{p}), \tag{25.1}$$

$$\dot{m}_c = -\rho_v\alpha_l(1-\alpha_l)\frac{1}{R}\psi_c\sqrt{\frac{2}{3}\frac{\bar{p}-p_{sat}}{\rho_l}} \quad (p_{sat} < \bar{p}), \tag{25.2}$$

Here, the dimensionless functions of growth $\psi_v$ and collapse $\psi_c$ are given by Eqns.16 and 24, respectively, their variations against bubble radius are plotted in Fig. 3. Noting that the smaller bubble has larger value during both dynamic processes. For growth period, tiny bubble possesses long expansion time ($\tau_v$) which is cubic proportional to volume change that leads to strong growth intensity. As for collapsing, the bubble speed at the direction of inward radial is accelerated gradually so that the intensity increases.

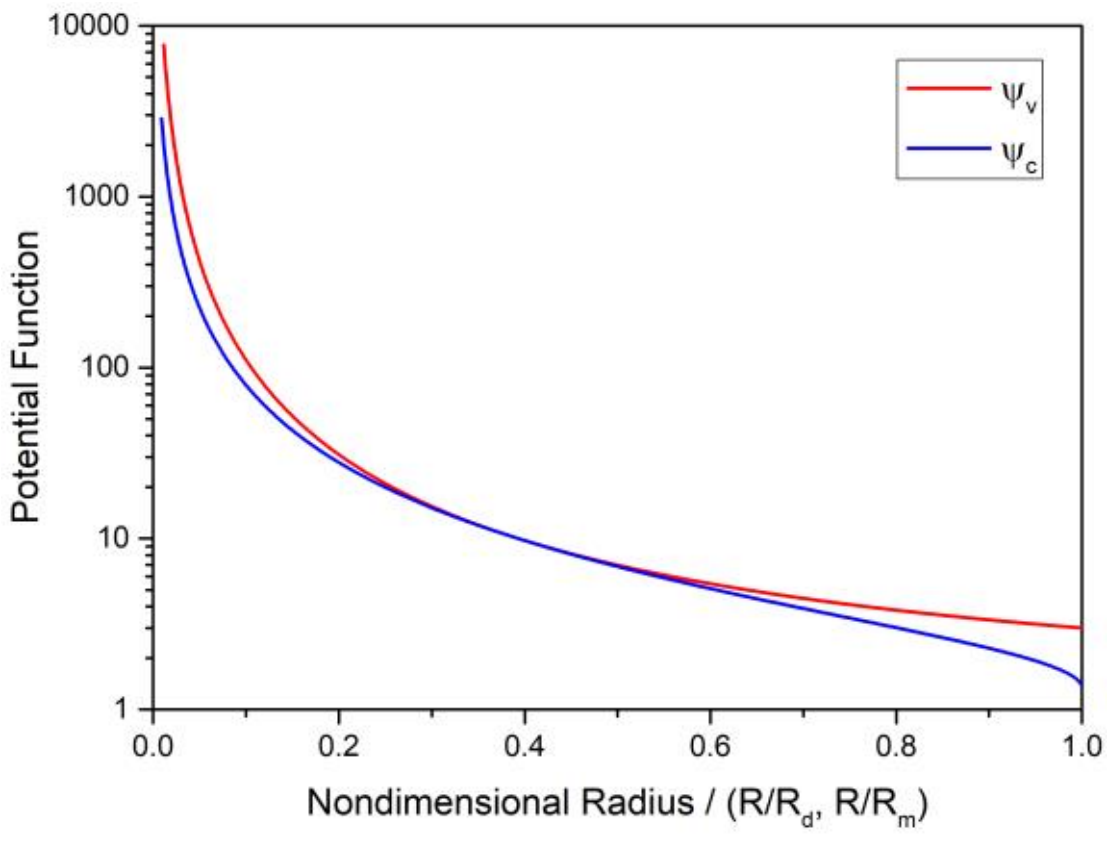

Fig. 3 The graphs of potential functions, $\psi_v$ and $\psi_c$

Apparently, only two parameters with significant physical meaning of cavitating bubbles, namely the Blake radius $R_b$ and the average maximum radius $R_m$, are employed to regulate the model application. The empirical coefficients are substituted by the nonuniform potential functions, $\psi_v$ and $\psi_c$, considered the nonlinear effects for both growth and collapse periods. Comparing with the Schnerr-Sauer model given in the following equations,

$$\dot{m}_v^S = C_v * \frac{3\rho_l\rho_v}{\rho}\alpha_l(1-\alpha_l)\frac{1}{R}\sqrt{\frac{2(p_{sat}-\bar{p})}{3\rho_l}} \quad (\bar{p} < p_{sat}), \tag{26.1}$$

$$\dot{m}_c^S = -C_c * \frac{3\rho_l\rho_v}{\rho}\alpha_l(1-\alpha_l)\frac{1}{R}\sqrt{\frac{2(\bar{p}-p_{sat})}{3\rho_l}} \quad (\bar{p} > p_{sat}). \tag{26.2}$$

the difference is, except the additional potential functions, that the density factor in front can be approximated as $\frac{\rho_l \rho_v}{\rho}\alpha_l \approx \rho_v$ based on $\rho_l \gg \rho_v$. The contribution of $\alpha_l$ vanishes that indicates both source terms are only proportional to $(1-\alpha_l)$ but not $\alpha_l(1-\alpha_l)$.

Moreover, the Schnerr-Sauer model can be reformulated as the same form of NDCM. In view of the symmetrical feature, we only discuss the collapse term as follow:

$$\dot{m}_c^S = -\rho_v \alpha_l (1-\alpha_l)\frac{1}{R}\psi_c^S \sqrt{\frac{2(\bar{p}-p_{sat})}{3\rho_l}} \quad (\bar{p} > p_{sat}). \tag{27.1}$$

where $\psi_c^S \cong 3C_c/\alpha_l$. Using the Eqn. 2 to substitute $\alpha_l$, we can get the equivalent collapse potential function of Schnerr model:

$$\psi_c^S = 3C_c\left(1+\frac{4}{3}\pi n R^3\right) \tag{27.2}$$

In contrast to $\psi_c$ which is inversely proportional to $R$, $\psi_c^S$ presents positive relation that will conclude a confusing interpretation that larger bubbles have stronger intensity during collapse. The comparison of $\psi_c$ and $\psi_c^S$ will be discussed in section 4.1.

# 3. Mathematical formulations and numerical method

## 3.1 Governing equations

The methodology in OpenFOAM for describing the two-phase system of cavitating flow is based on Eulerian's homogenous mixture approach, in which both phases are treated as incompressible, isothermal, and immiscible. It is simple and efficient to employ the one-field formulation (OFF) of Navier-Stokes equations that the properties of two phases, including density and viscosity, are hybrid as equivalent single-phase flow. The filtered governing equations including transport equation of liquid volume fraction and momentum conservation of effective fluid are given in Eqn. 28 and 29 respectively, and the velocity divergence in Eqn. 30 comes from summing over the volume fraction equations of liquid and vapor that enables to build the pressure equation to update the flux field. More details can be referred to Fleckenstein [6] for a rigorous derivation:

$$\frac{\partial \alpha_l}{\partial t} + \boldsymbol{U}\cdot\nabla\alpha_l = \frac{\rho}{\rho_l\rho_v}(|\dot{m}_c| - |\dot{m}_v|), \tag{28}$$

$$\frac{\partial}{\partial t}(\rho\boldsymbol{U}) + \nabla\cdot(\rho\boldsymbol{U}\boldsymbol{U}) = -\nabla p_{rgh} + \nabla\cdot[\mu(\nabla\boldsymbol{U}+\nabla^T\boldsymbol{U}) - \boldsymbol{\tau}_T] - \boldsymbol{g}\cdot\boldsymbol{h}\nabla\rho, \tag{29}$$

$$\nabla\cdot\boldsymbol{U} = \left(\frac{1}{\rho_l} - \frac{1}{\rho_v}\right)(|\dot{m}_c| - |\dot{m}_v|). \tag{30}$$

Here, two phases are assumed to share the same velocity denoted as $\boldsymbol{U}$. The hybrid density $\rho$ and viscosity $\mu$ are weighted based on volume fraction $\alpha_l$ linearly with the constant properties of each phase, namely $\rho_l$, $\rho_v$, $\mu_l$, $\mu_v$, which are:

$$\rho = \rho_l \alpha_l + \rho_v \alpha_v, \tag{31.1}$$
$$\mu = \mu_l \alpha_l + \mu_v \alpha_v. \tag{31.2}$$

Note that the pressure $p_{rgh}$ is relative to hydrostatic pressure $\rho \boldsymbol{g} \cdot \boldsymbol{h}$ to avoid the algorithmic trouble of artificial diffusion induced by height difference. The non-linear stress $\boldsymbol{\tau}_T$ is closed by RANS or LES turbulence models. In the present work, we employed the k-omega-SST two equation model [40]. Thus, the complete framework is composed of three parts which are multiphase model (Eqn.28, 29 and 30), turbulence model ($\boldsymbol{\tau}_T$ in Eqn.29) and cavitation model ($\dot{m}_c$ and $\dot{m}_c$ in Eqn.28).

## 3.2 Algorithms and discretization

The above governing equations are implemented into CFD code based on the solver of interPhaseChangeFoam, in which two important algorithms are developed for capturing the topological changes of bubble cluster and coupling the velocity-pressure to prevent checkerboard distribution.

One of the mature algorithms to solve a convective-only transport equation is VoF-based interface capturing method that a new high-resolution algebraic reconstruction proposed by Weller [34] based on flux-corrected transport (FCT) is implemented in OpenFOAM. For further sharpness, comparing to the traditional approaches of compressive schemes like HRIC or CICSAM, the "counter-gradient" diffusion term is joined into transport equation to compress the interface in the reverse direction of volume fraction gradient that has good performance for 2D and 3D complex flows. It should be noted that the interface of cavitation cavity refers to the boundary of vapor-liquid mixture ($0 < \alpha_l < 1$) and pure liquid ($\alpha_l = 1$) differently from the one between water ($\alpha_l = 1$) and vapor ($\alpha_l = 0$) where the ambiguous region $(0 < \alpha_l < 1)$ should be contracted as sharper as possible. Therefore, it is improper to employ the compression term which would bring unreasonable diffusion inside cavity. We recommend to use the FCT-based numerical method of semi-implicit multi-dimensional limiter for explicit solution (MULES) for better boundedness and consistency. A detailed description of this algorithm can be found in [33].

The SIMPLE/PISO algorithm on collocated grid is realized via the technique of momentum interpolation proposed by Rhie-Chow [41] that the serrated pressure would be eliminated by introducing the third pressure derivative in correction equation. However, only an incomplete method is implemented in OpenFOAM for robustness. The simulation results would be depending on time step size so that we suggest a modified momentum interpolation (MMI) method referring to the work by Cubero [35, 36] to remove the drawback. The original momentum interpolation (OMI) in OpenFOAM is given by:

$$\boldsymbol{U}_f^* = \overline{\frac{1}{a_P}\boldsymbol{H}^*} - \overline{\frac{1}{a_P}}\nabla p_{rgh,f}^* + \epsilon * \overline{\frac{1}{a_P}}\,\overline{a_P^t}\left(\boldsymbol{U}_f^n - \overline{\boldsymbol{U}_P^n}\right), \tag{32}$$

that the operator of hat bar is linear interpolation from cell to face center, and the superscript $*$ means the mid-iteration till convergence to $n+1$. The discrete coefficient $a_P$ can be decomposed into temporal $a_P^t$ and spacial $a_P^s$ dependent items, and the vector $\boldsymbol{H}$ represents the collection of neighbor points summation and other sources. Noting that the last term, called Choi correction, is emerged in unsteady problems that the flux difference of previous time step is used to correct the interpolated velocity, where $\epsilon$ employed in OpenFOAM is the empirical factor less than unity to prevent the correction value inducing instability. Though, theoretically, the Choi correction would be vanished through several iterations, the time step is contained in coefficient $a_P$ that leads to the convergence result still associating with time. To remedy the problem, the specific value is defined as $d_P = a_P^t / a_P^s$, thereby the Rhie-Choi interpolation can be reformulated as:

$$\boldsymbol{U}_f^* = \overline{\frac{1}{1+d_P}\frac{1}{a_P^s}\boldsymbol{H}^*} - \frac{1}{1+d_f}\frac{1}{a_f^s}\nabla p_{rgh,f}^* + \left(\frac{1}{1+d_f}\frac{a_f^t}{a_f^s}\boldsymbol{U}_f^n - \overline{\frac{1}{1+d_P}\frac{a_P^t}{a_P^s}\boldsymbol{U}_P^n}\right). \tag{33}$$

Here, it is assumed that all the coefficients on face center are interpolated linearly by cell values, and the approximate relations are introduced:

$$\overline{\frac{1}{1+d_P}\frac{1}{a_P^s}\boldsymbol{H}^*} \approx \frac{1}{1+\overline{d_P}}\overline{\frac{1}{a_P^s}\boldsymbol{H}^*}, \tag{34.1}$$

$$\overline{\frac{1}{1+d_P}\frac{a_P^t}{a_P^s}\boldsymbol{U}_P^n} \approx \frac{1}{1+\overline{d_P}}\overline{d_P\boldsymbol{U}_P^n}. \tag{34.2}$$

Thus, the formulation of MMI can be derived by substituting Eqn34.1 and 34.2 into Eqn.33:

$$\boldsymbol{U}_f^* \cong \frac{1}{1+\overline{d_P}}\left(\overline{\frac{1}{a_P^s}\boldsymbol{H}^*} - \overline{\frac{1}{a_P^s}}\nabla p_{rgh,f}^*\right) + \frac{1}{1+\overline{d_P}}\left[\overline{d_P}\boldsymbol{U}_f^n - \overline{d_P\boldsymbol{U}_P^n}\right]. \tag{35}$$

To ensure stability of solving pressure equation, the cavitation source term in Eqn.30 is handled as semi-implicit form based on the principal diagonal dominant as follows:

$$\nabla\cdot\left[\frac{1}{1+\overline{d_P}}\left(\overline{\frac{1}{a_P^s}\boldsymbol{H}^*} - \overline{\frac{1}{a_P^s}}\nabla p_{rgh,f}^*\right) + \frac{1}{1+\overline{d_P}}\left[\overline{d_P}\boldsymbol{U}_f^n - \overline{d_P\boldsymbol{U}_P^n}\right]\right]$$
$$= \left(\frac{1}{\rho_l} - \frac{1}{\rho_v}\right)F(\alpha_l)\sqrt{\frac{2}{3\rho_l(|p_{sat} - \bar{p}^n| + 0.001 * p_{sat})}}\left[p_{rgh}^* - (p_{sat} - \rho\boldsymbol{g}\cdot\boldsymbol{h})\right], \tag{36.1}$$

where $F(\alpha_l)$ is $\alpha_l$ dependent function:

$$F(\alpha_l) = \rho_v\alpha_l(1-\alpha_l)\frac{1}{R}\left[\psi_v * pos(p_{sat} - \bar{p}) + \psi_c * neg(p_{sat} - \bar{p})\right]. \tag{36.2}$$

The Choi correction will disappear during PISO/SIMPLE loops, and the time step is excluded out of $a_P^s$. We employ two cases of lid-driven cavity (single phase flow) and 2D bubble rising [42] (multiphase flow) to test the performance of different methods shown in Fig. 4. It is seen that the results obtained by MMI are overlapped completely to indicate the independence of time step (Fig. 4 (b) and (d)) for both situations.

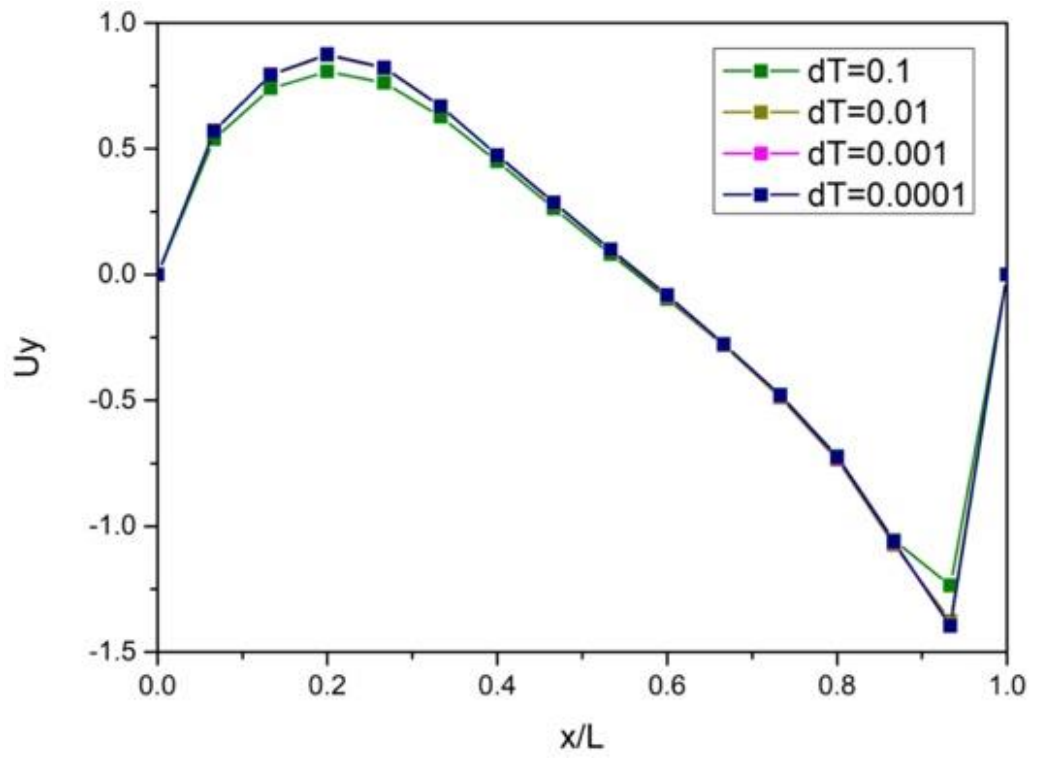

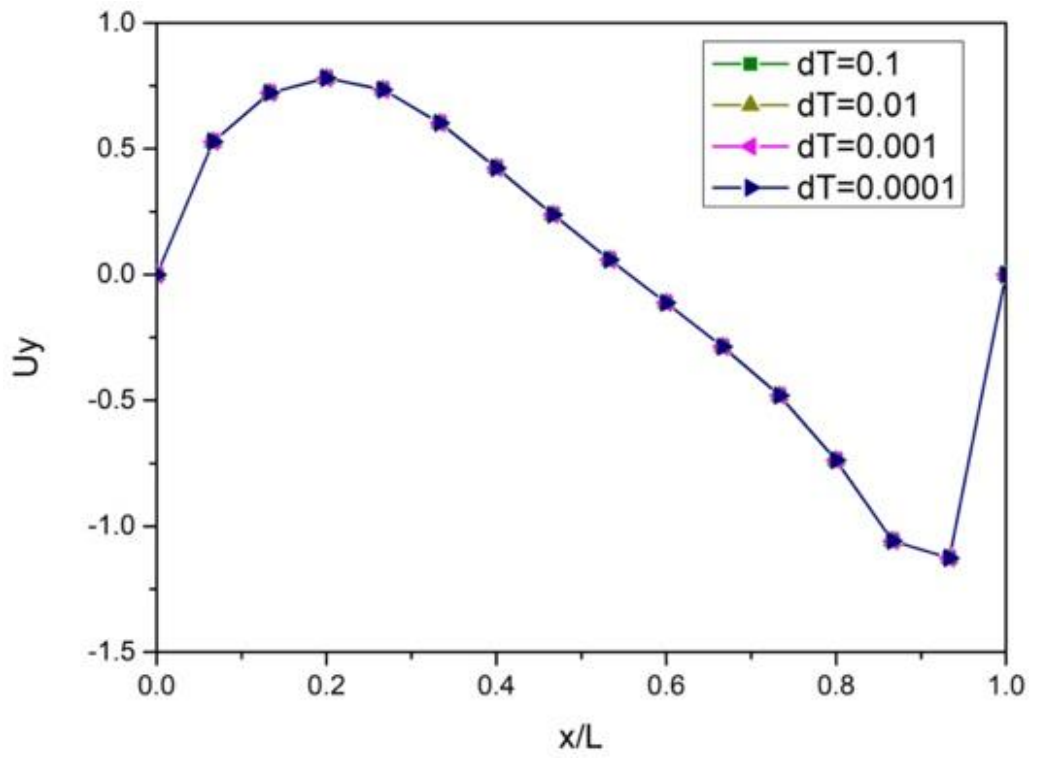

(a) Lid-driven cavity (OMI) (b) Lid-driven cavity (MMI)

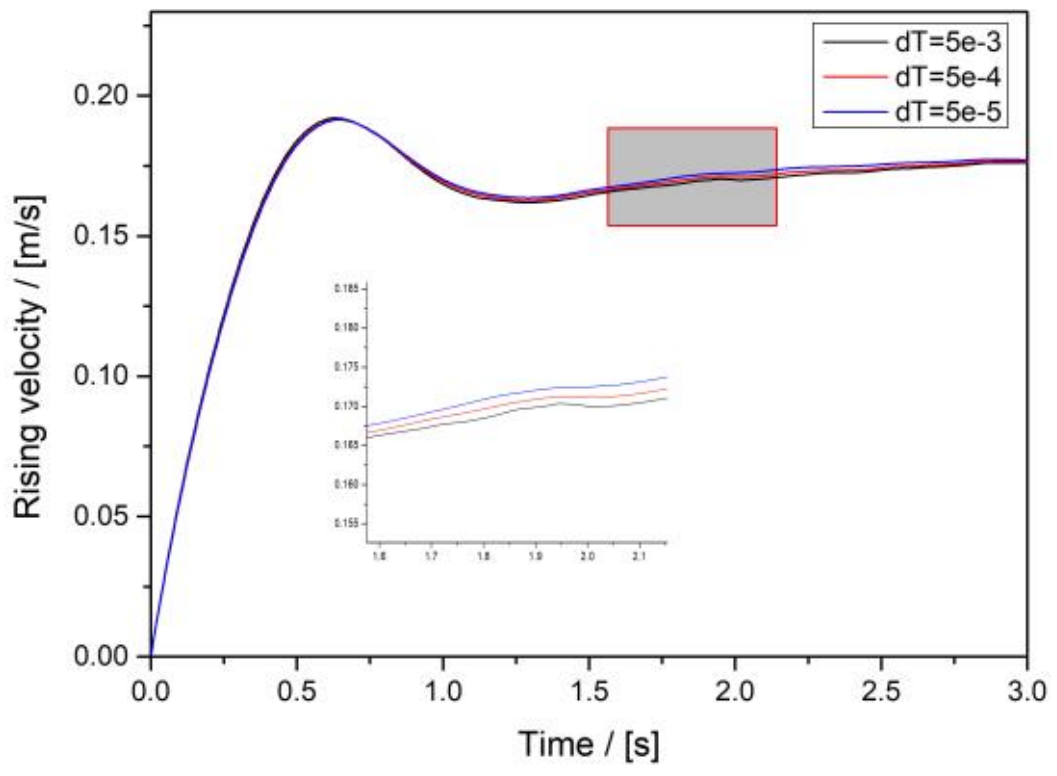

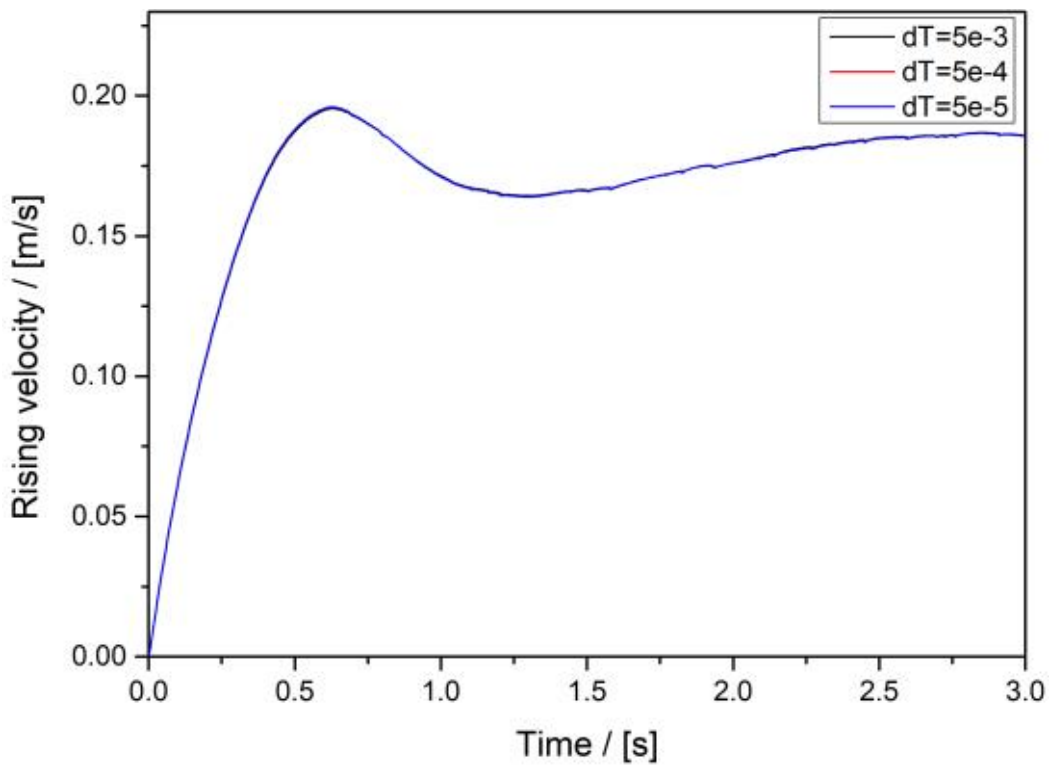

(c) Rising bubble (OMI) (d) Rising bubble (MMI)

Fig. 4 Comparison of the velocity profiles between OMI and MMI against different time steps

## 3.3 Numerical configuration

The strategy of velocity-pressure coupling is designed dual loops as called PIMPLE which have inner PISO and outer SIMPLE for accommodating large time step size [28]. The TVD type high order resolution schemes are used for the convective terms, and the second order central difference scheme is used for the diffusion terms. The first-order implicit Euler scheme is used for the transient terms.

# 4. Validation cases

In this section, the main purpose is to validate the cavitation model (NDCM) developed in this work through three fundamental cases. For lack of exact solution about cavitating flows, the first case is to simulate the collapse process of vapor bubble cluster to illustrate differences between linear and nonlinear models. The real bubble cluster excited by ultrasonic field is then investigated to reveal physical connotation of model parameters through a series of simulations for different experimental conditions. The new model is finally applied to the convective bubble cloud in hydrodynamic cavitation of slender bodies with conical or blunt nose to further highlight the capability of NDCM.

## 4.1 Bubble cluster collapse

The necessity of considering the nonlinear effects represented by the potential function of $\psi_c$ during bubble collapse can be demonstrated clearly by this validation case. Here, we recommend to use the simulation case by Schmidt [29] who developed a thermodynamic equilibrium model that the interface can be resolved implicitly when the grid resolution is sufficiently fine so as comparable to DNS. As is shown in Fig. 5(a), the bubble cluster above the wall covers a spherical domain with a diameter of $r_b = 30\text{mm}$ within which $N = 150$ spherical bubbles of equal radii $r_0$ distributed from dense center to sparse border are randomly generated. All bubbles are filled with water vapor while the surrounding domain contains liquid water at an initial pressure of $p_\infty = 100bar$. The initial pressure inside the bubbles is equal to the vapor pressure $p_{sat} = 2340Pa$. The velocity field is initially at rest.

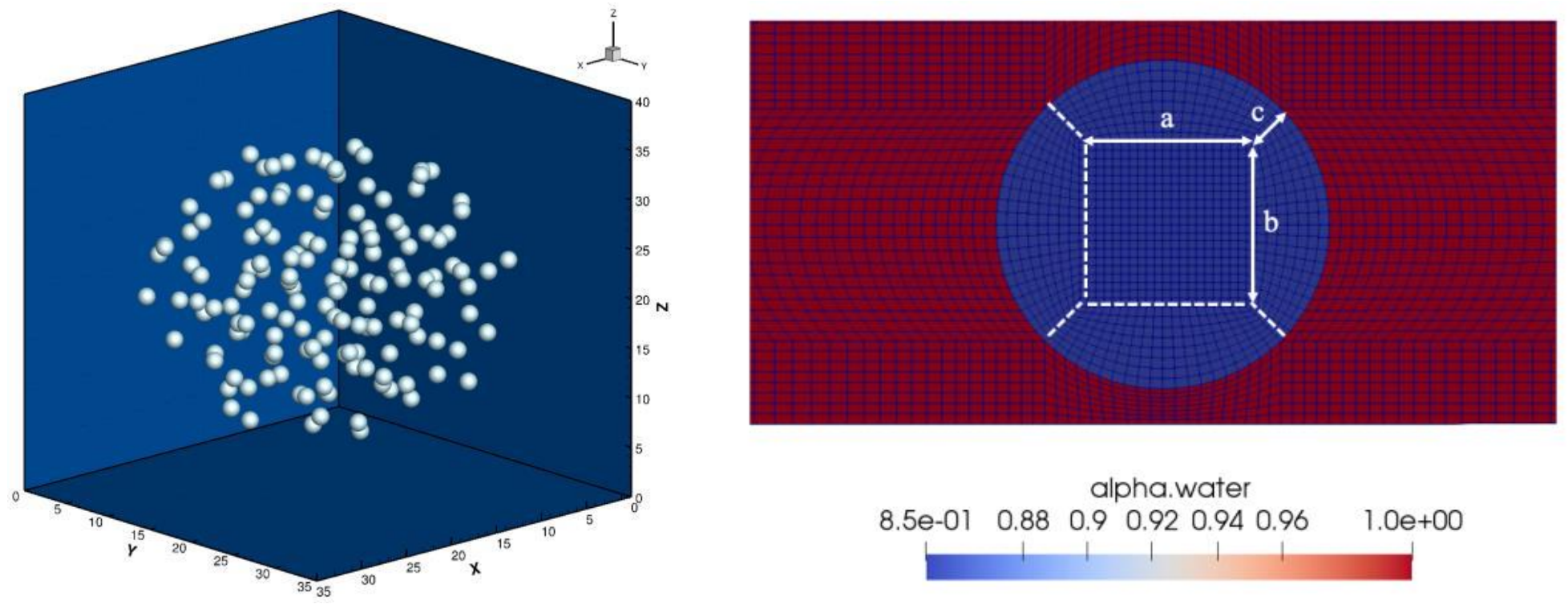

(a) (b)

Fig. 5 Distribution of bubble cluster (a), lateral section of computational mesh (b)

The simulation is carried out for bubbles with the radius $r_0$ of 1mm and 1.5mm by NDCM and Schnerr-Sauer models. The lateral section of 3D hexahedral grid is shown in Fig. 5(b), three mesh resolutions for the spherical domain containing under-resolved bubbles are tested by the nodes of $a \times b \times c$. We introduce a factor $\lambda$ that is defined as the ratio between bubble radius $(r_0)$ and the numerical resolution scale $(\sqrt[3]{\bar{V}})$, where $(\sqrt[3]{\bar{V}})$ is the cubic root of cell-weighted average volume in bubble cluster. The model parameters and initial volume fraction $\alpha_l$ can be calculated from:

$$\alpha_l = 1 - N * \left(\frac{r_0}{r_b}\right)^3, \tag{37.1}$$

$$n = \frac{1-\alpha_l}{\alpha_l}\frac{1}{\frac{4}{3}\pi r_0^3}, \tag{37.2}$$

that the setup details are shown in the Table 1. According to the conclusions from the reference paper, the bubble distribution influences the pressure field significantly but trivial for the collapse period, so we initialize the uniform field for $\alpha_l$ to quantitatively investigate the main collapse process whereas only compare the pressure results qualitatively. The numerical pressure transducer is used consistently with [29] that the sampling frequency is 49.3MHz on the wall of $15 \times 15mm^2$ directly beneath the bubble cluster.

Table 1 Initializations for the vapor bubble cluster

| | | $r_0 = 1.0\ [mm]$ | $r_0 = 1.5\ [mm]$ |
|---|---|---|---|
| Liquid volume fraction $\alpha_l\ [-]$ | | 0.95556 | 0.85 |
| Bubble number density $n\ [1/m^3]$ | | 11103833 | 12482740 |
| Resolution ratio $\lambda = \frac{r_0}{\sqrt[3]{\bar{V}}}\ [-]$ | $Mesh$ 1 (6.5k cells) | 0.386 | 0.579 |
| | $Mesh$ 2 (16k cells) | 0.521 | 0.782 |
| | $Mesh$ 3 (26k cells) | 0.613 | 0.919 |

The validation for grid independence is shown in Fig. 6, that all the variables are nondimensionalized by the equivalent radius $R_{eqn}$ and its Rayleigh time $\tau$. It can be seen that the collapse time calculated by NDCM (solid lines) is converged on Mesh 2 and 3 for both radii. Recalling the assumption III which is indicated that the bubble interactions could be ignored above the resolution of Mesh 2. Differently, the results by Schnerr model (symbol-solid lines) given the value of $C_c$ as unity are overlapped together which duration is longer than the former.

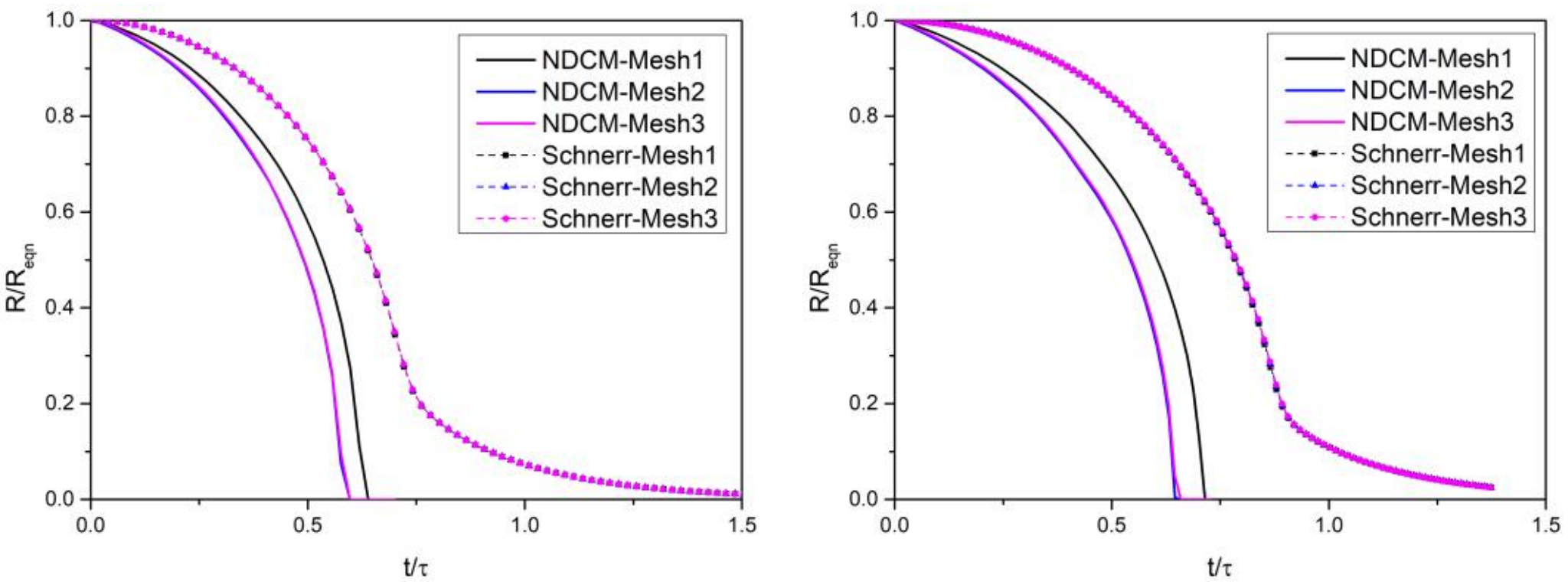

$r_0 = 1.0\ [mm]$ $r_0 = 1.5\ [mm]$

Fig. 6 The grid independence verifications for different meshes

Fig. 7 shows model comparisons on Mesh 2. The collapse time in reference paper is taken as benchmarks which accounts for 60% and 65% of Rayleigh time $\tau$ of the radius of 1.0mm and 1.5mm respectively. The collapse time interval by NDCM agrees well with the benchmarks although the rate of change has discrepancy which is likely due to employ the uniform distribution of $\alpha_l$ inconsistently with the original method. However, it is deviated widely by Schnerr model with $C_c = 1$, renamed as Schnerr-Cc as follows, which has insufficient intensity of source term that induces bubbles hard to collapse especially in the last stage. In our opinion, the problem is caused by model error that different potential functions of Eqns. 24 and 27.2 depict contradictory trend against bubble radius. It can be considered that there is a collapsing shell at the border of bubble cloud based on the fact that its character of collapse processes is layer-by-layer. The internal bubbles are normally larger than the external along the shell radial. The potential function $\psi_c^S$ in Schnerr model gives a positive correlation with bubble radius that may induce the high intensity of source term to be emerged inside bubble cloud. Thus, the external tiny bubbles are constrained by weak sources that leads to difficult collapse. Even though $\psi_c^S$ is incompatible with physical actuality, the Schnerr model can still be used by changing the coefficient $C_c$. Ghahramani et al. [43] studied the Schnerr approach to simulate bubble cluster collapse, and they employed a high $C_c$ to obtain better results of collapse period but emerged huge numerical pressure wiggles unexpectedly. We also tried larger values $C_c$ of 800 and 1200 for both cases that make the collapse period matched well with the benchmarks. As is shown in Fig. 8, highest pressure pulse occurs at the last process of collapse. However, only the result by Schnerr-1 has smooth curve whereas others do not. Comparison of the results by NDCM (red line) and Schnerr with large $C_c$ (blue line), we can see that the new model can suppress most extents of spurious pressure wiggles. Noting that the only difference in the contrast is the potential functions that indicate the physics implied by $\psi_c$ is more reasonable than $\psi_c^S$. Comparison of the influence by different meshes (red and green lines) of NDCM results show that finer resolution contributes to control unphysical oscillations, particularly in $r_0 = 1.5$. It can be found that the pressure is more sensitive to mesh resolution although the collapse period has been converged.

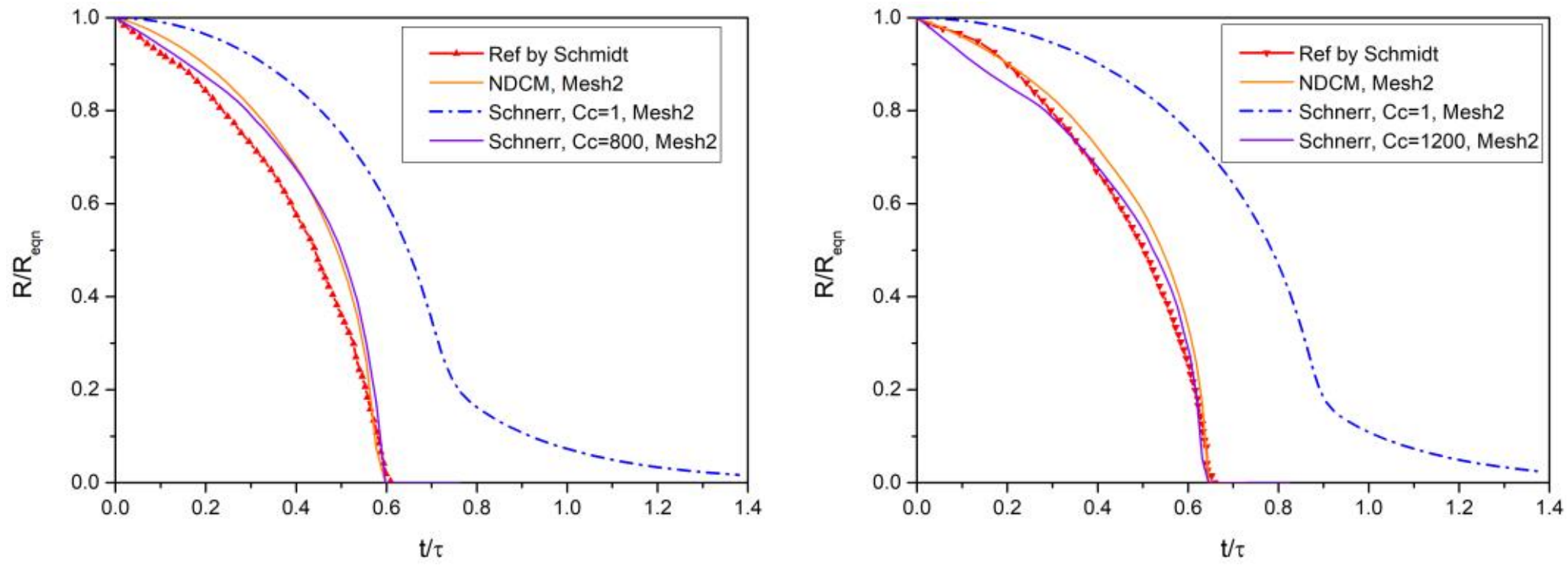

(a) $r_0 = 1.0\ [mm]$ (b) $r_0 = 1.5\ [mm]$

Fig. 7 Comparison of time history of bubble cluster collapse

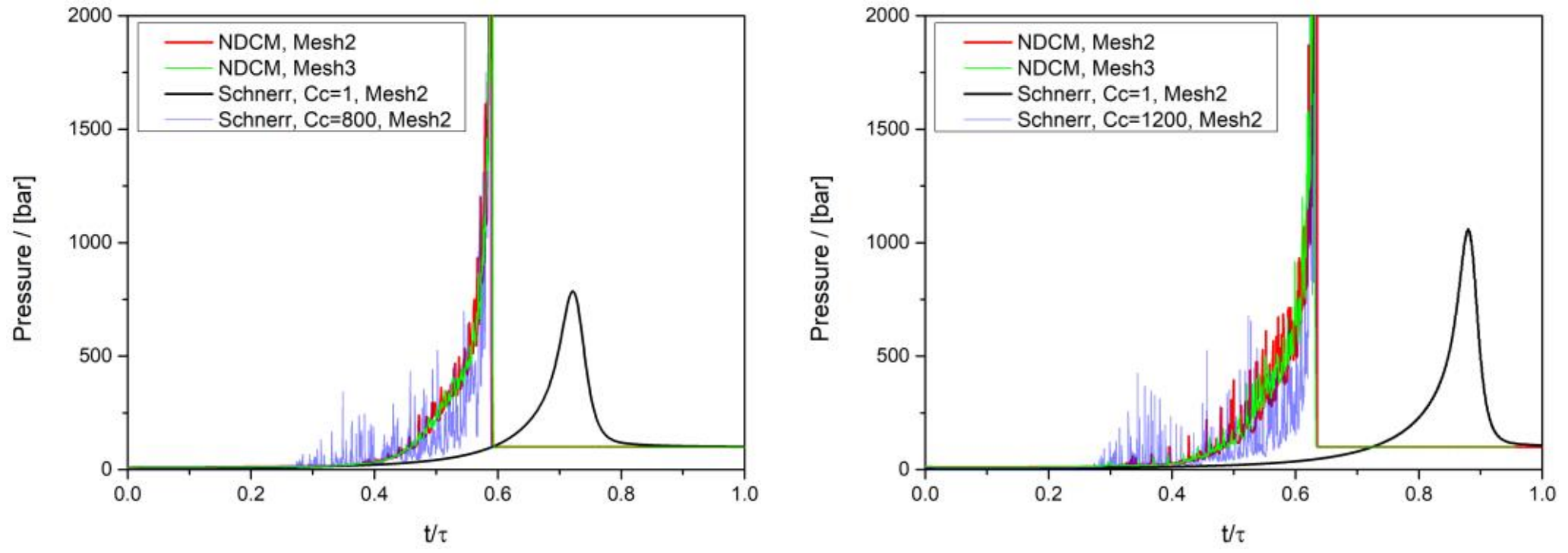

(a) $r_0 = 1.0\ [mm]$ (b) $r_0 = 1.5\ [mm]$

Fig. 8 Comparison of time history of pressure transducer on the wall

For further comparison and understanding the differences between two models, the bubble cloud structure, pressure contours and streamlines for the case of $r_0 = 1mm$ at different time instances are shown in Fig. 9. The behaviour of bubble cluster collapse is seen in the left sides that the variation of cloud radius is tardy from $t = 0$ to $t = 0.618T$, subsequently dramatic collapse occurs during the rest of times. As is mentioned in assumption III, interactions inside bubble cloud can be reflected in the flux transfer between neighbor cells even if the local interactions are ignored. The motivation is that the velocity doesn't be divergence free (Eqn.30) that indicating the velocity field will be affected by cavitation source terms which inverse gradient will generate streamlines in the same direction. As is seen from left sides in Fig. 9(b)~(d), the structures of velocity field of NDCM and Schnerr-1 are quite similar. The high-speed collapsing shell is driven by appropriate source term where large values distribute at the outside, but the low-speed internal region almost immunes to collapse until the last period at $t = 0.927T$. However, the situation in Schnerr-800 is that unreasonable inverse flow is formed due to the higher source values emerge inside that sometimes induces the internal bubbles collapse priorly instead of the external (red dash circle in Fig. 9(b. II)). The temporal evolution of bubble radius distribution in the direction of $\rho_r$ is shown in Fig. 10. The serrated line marked by the dash circle at $t = 0.618T$ illustrates the bubble collapsing sequence is from inside-out. Moreover, the innermost bubbles are affected prematurely from the boundary (dash arrow).

Apparently, the property of potential functions ($\psi_c$ and $\psi_c^S$) is the main reason to influence the source term distribution. Besides, the local pressure $\bar{p}$ (in Eqns. 26.2 and 27.1) is another key factor that exhibits positive correlation to source terms, so that high pressure around outer bubble cluster enhances local sources. It should be mentioned that $\bar{p}$ is compatible with $\psi_c$ that will further strengthen outer source values, but $\psi_c^S$ weakens the effect by $\bar{p}$. For the case of Schnerr-1, it can be inferred that $\psi_c^S$ is trivial relative to pressure $\bar{p}$ which dominates the source term over the whole collapse period, thus large source values only appear externally to avoid inverse flows. However, such distribution of source term is destroyed by employing large $C_c$ which is augmented 800 times.

It is seen from right sides in Fig. 9 that there are some local pressure pulses around bubble cloud for contours of NDCM and Schnerr-800. These pulses are the reasons for wiggles detected by the wall pressure transducer. Rossinelli et al. [44] has implemented a two-phase flow DNS to simulate 15000 bubbles collapse, which found that the pressure wiggles should be existed during temporal evolution. Thus, the smooth pressure profile predicted by Schnerr-1 is unreasonable. In Fig. 9(b.III)~(e.III), the bubble cloud is dispersed excessively due to insufficient intensity of cavitation sources that the dissipated pressure gradient prevents local high pressure to happen. Comparison of pressure field by two models in Fig. 9(d. I) and (d. II), the positions of high pressure predicted by NDCM are located at the external shell from where to infinity the values decrease monotonously to ambient pressure. However, the corresponding points in Schnerr-800 invade into the bubble cloud where pressure pulse is surrounded by an additional low-pressure band (red dash arrow). We suppose that the misplacing pressure distribution is one of the reasons to cause the spurious pressure pulses which can be eliminate effectively by using the derived potential function $\psi_c$ in NDCM.

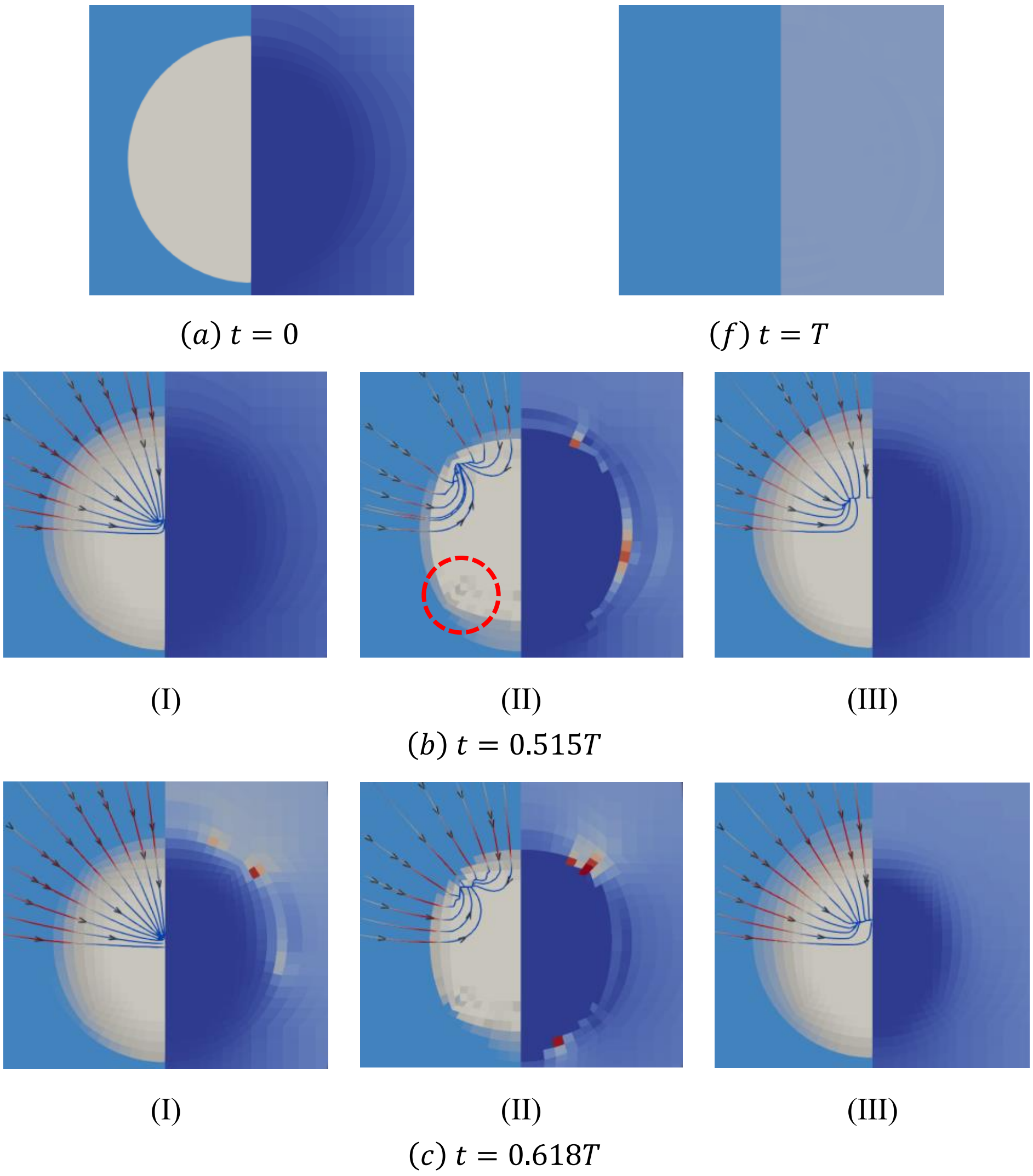

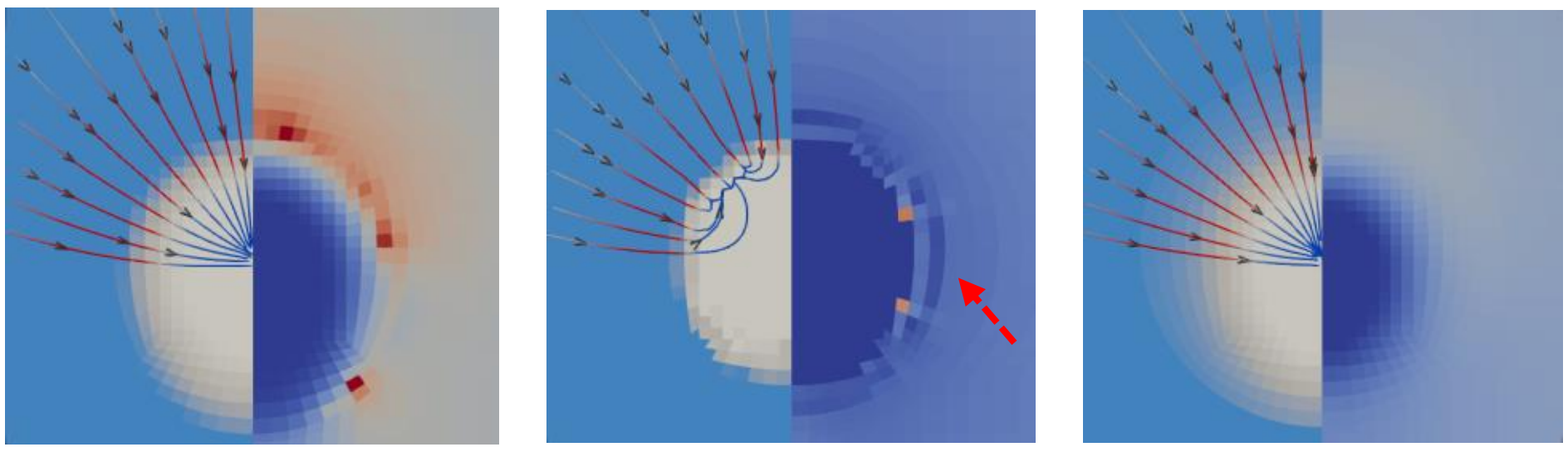

(I) (II) (III)

$(d)\ t = 0.755T$

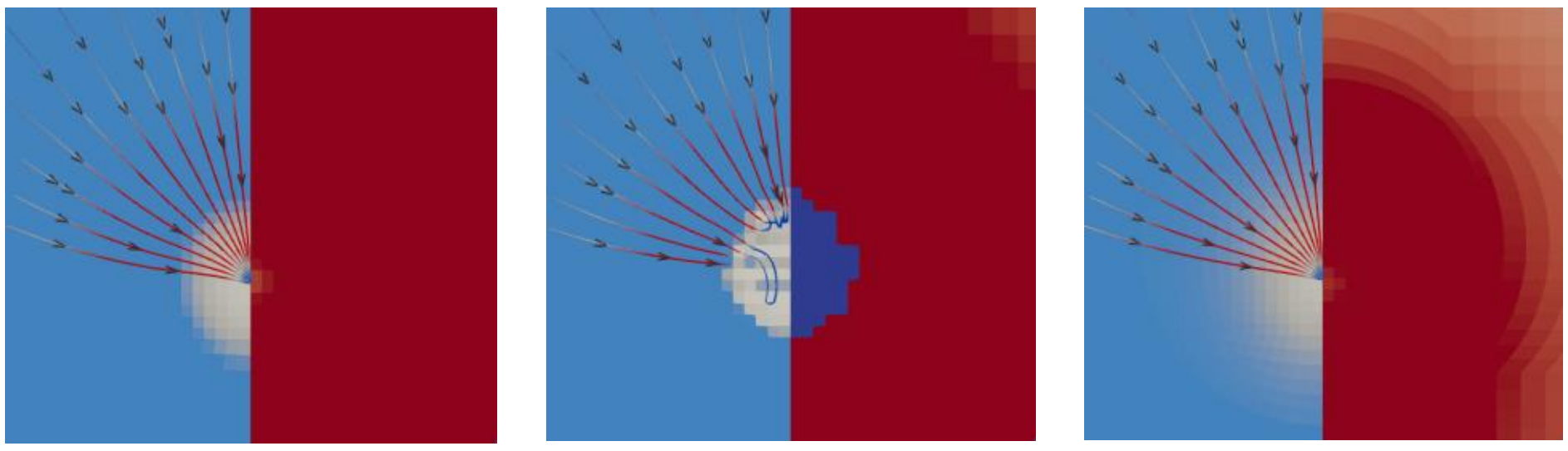

(I) (II) (III)

$(e)\ t = 0.927T$

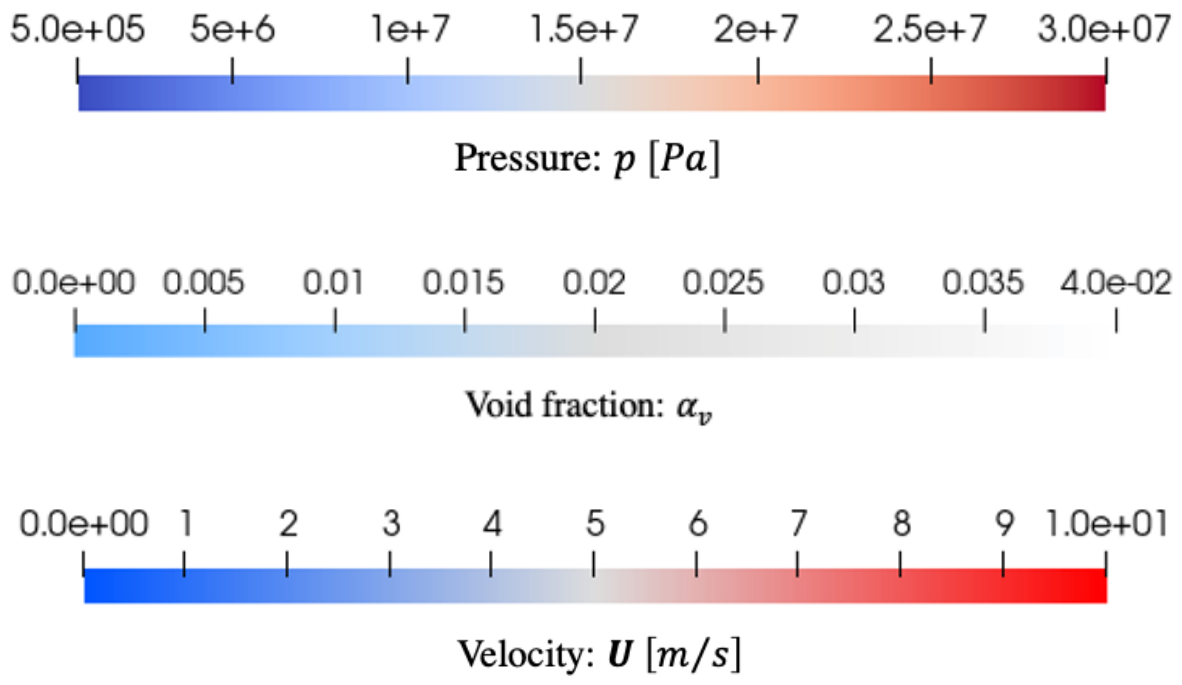

Fig. 9 Comparison of NDCM (I), Schnerr with $C_c = 800$ (II), and Schnerr with $C_c = 1$ (III) in prediction of cloud structure (left side), pressure contours (right side) and streamlines at different time instances

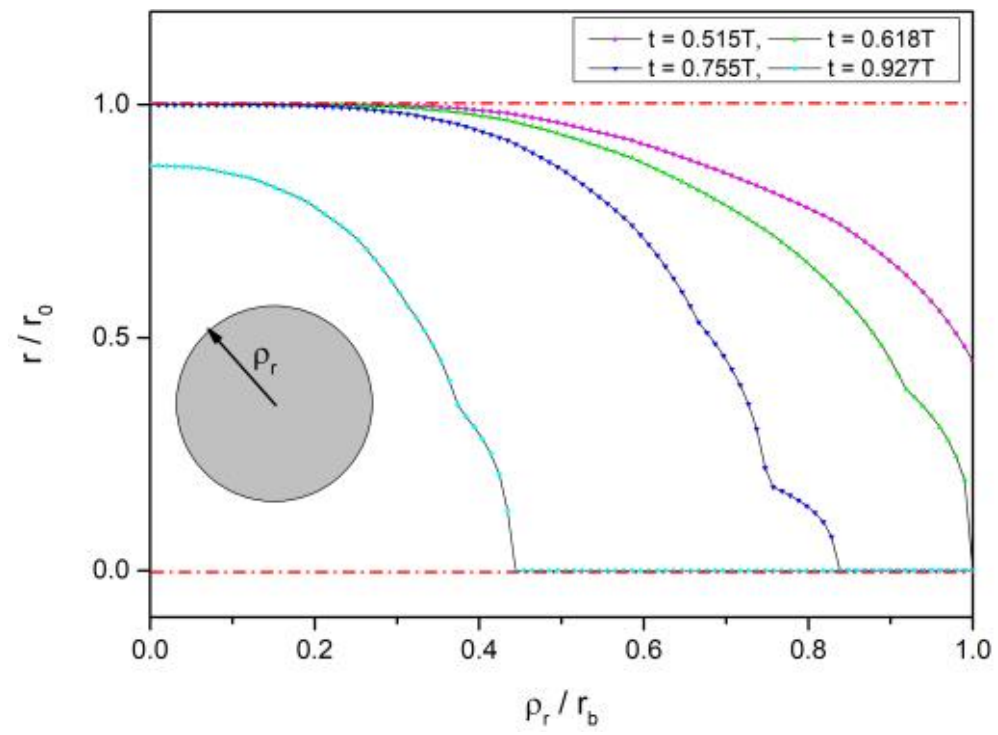

(a) NDCM

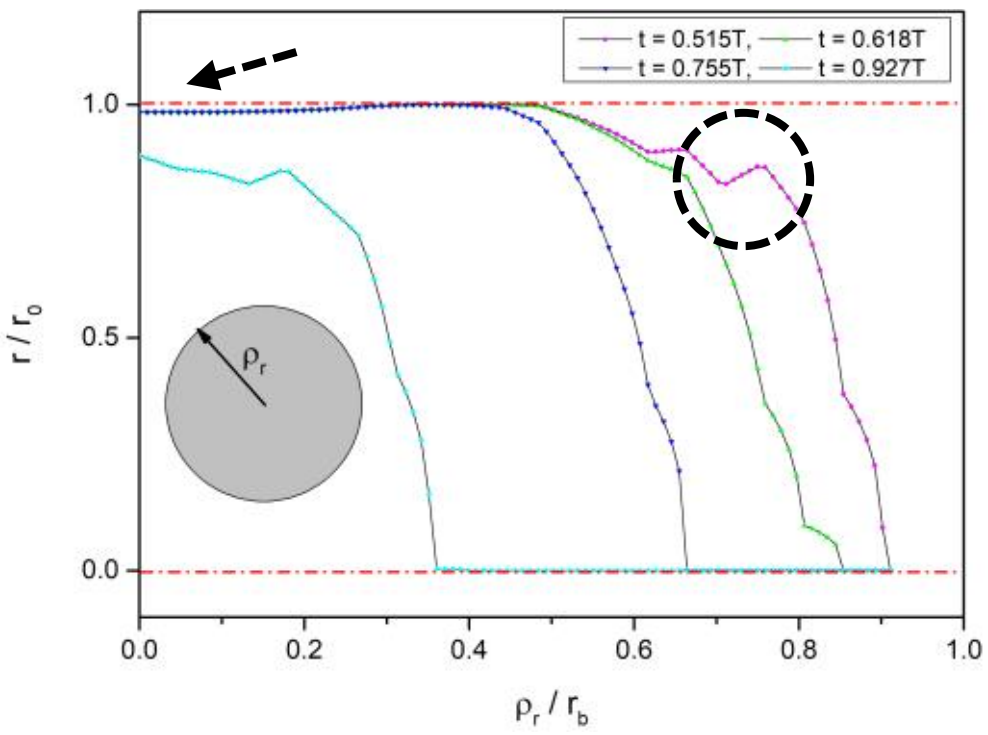

(b) Schnerr $(C_c = 800)$

Fig. 10 Comparison of distribution of bubble radius in radial direction of bubble cluster

## 4.2 Ultrasonic horn

After exhibiting the basic expectation of NDCM, the discussion will be concentrated on the real bubble cloud generated by the high frequency oscillating ultrasonic horn. It has been observed that if the horn tip is sufficiently small and driven at high amplitude, cavitation is very strong and the tip can be covered entirely by the gas/vapor phase for longer time intervals. We employ the experiment designed by Žnidarčič et al. [30] who investigated a systematic study in water at a 20 kHz with the horn diameter of 3 mm under variation of driving power, air saturation, viscosity, surface tension and temperature, that the attached cavity emerged peculiar dynamics with a self-generated frequency of expansion and collapse periodically. After that, they [31] carried out the correspondingly simulation studies and obtained poor predictions of flow features with the original TEM method, i.e., Schnerr-Sauer model. In their opinion, the Schnerr-Sauer-like model cannot adapt to the rapidly changing driving pressures, they presented an improved approach which also considered the second derivative term of Rayleigh-Plesset equation but in differential form. Good agreements comparing with measurements were then revealed for cavity shape and its frequency. However, the evolutionary tendency of the bubble cloud does not match well with experiment especially in expansion process. In this section, the tasks are not only to demonstrate better predictive results by applying the NDCM method but also the rules of parameter regulation. Table 2 lists five experimental conditions for water at room temperature by which comparison of simulations can validate the physical meaning of model parameters.

Table 2 Four experimental conditions for ultrasonic horn

| Case | Percentage of max power [%] | Vibrating amplitude $A_h$ [$\mu m$] | Saturation [%] |
|---|---|---|---|
| A | 70 | 164 | 100 |
| B | | | 50 |
| C | | | 20 |
| D | 30 | 100 | 100 |

### 4.2.1 Model parameters determination

Recalling that the model parameters, $R_b$ and $R_m$, are the particular points living on the bubble dynamics curve, it is therefore demanded to determine the initial bubble state to estimate their reasonable values. The theory of rectified mass diffusion which is the mechanism of cavitation inception in acoustic field is adopted to describe the formation of microbubbles from dissolved gas. Here, the bubble nuclei are formed from microbubbles by gradually mass transfer, between which the difference of magnitude order is commonly at one. The analytical model for air-water systems proposed by Crum [45] that there exists a certain critical amplitude of acoustic pressure $\tilde{p}_c$ above which the microbubbles will begin to grow by rectified diffusion. The expression is given as follow:

$$\tilde{p}_c = \rho R_0^2 \omega_N^2 \sqrt{\frac{\left[\left(1-\frac{\omega^2}{\omega_N^2}\right)^2 + \left(\frac{b\omega}{\omega_N}\right)^2\right]\left(1+\frac{2\sigma}{R_0 p_\infty} - \frac{c_i}{c_0}\right)}{(3+4K)\left(\frac{c_i}{c_0}\right) - \left[\frac{3(\eta-1)(3\eta-4)}{4} + (4-3\eta)K\right]\left(1+\frac{2\sigma}{R_0 p_\infty}\right)}}, \quad (38)$$

where the $R_0$ is the microbubble radius, the $c_i$ and $c_0$ are the concentration at bubble interface and ambient liquid respectively, which ratio represents the gas saturation, whereas other variables can be seen in [45]. Shown in Fig. 11 are the values of Eqn. 38 for the driving frequency at 20kHz. It is indicated that large microbubbles tend to grow easily on the same saturation curve, besides, it is difficult to form nuclei for the same size of microbubble in degassed water.

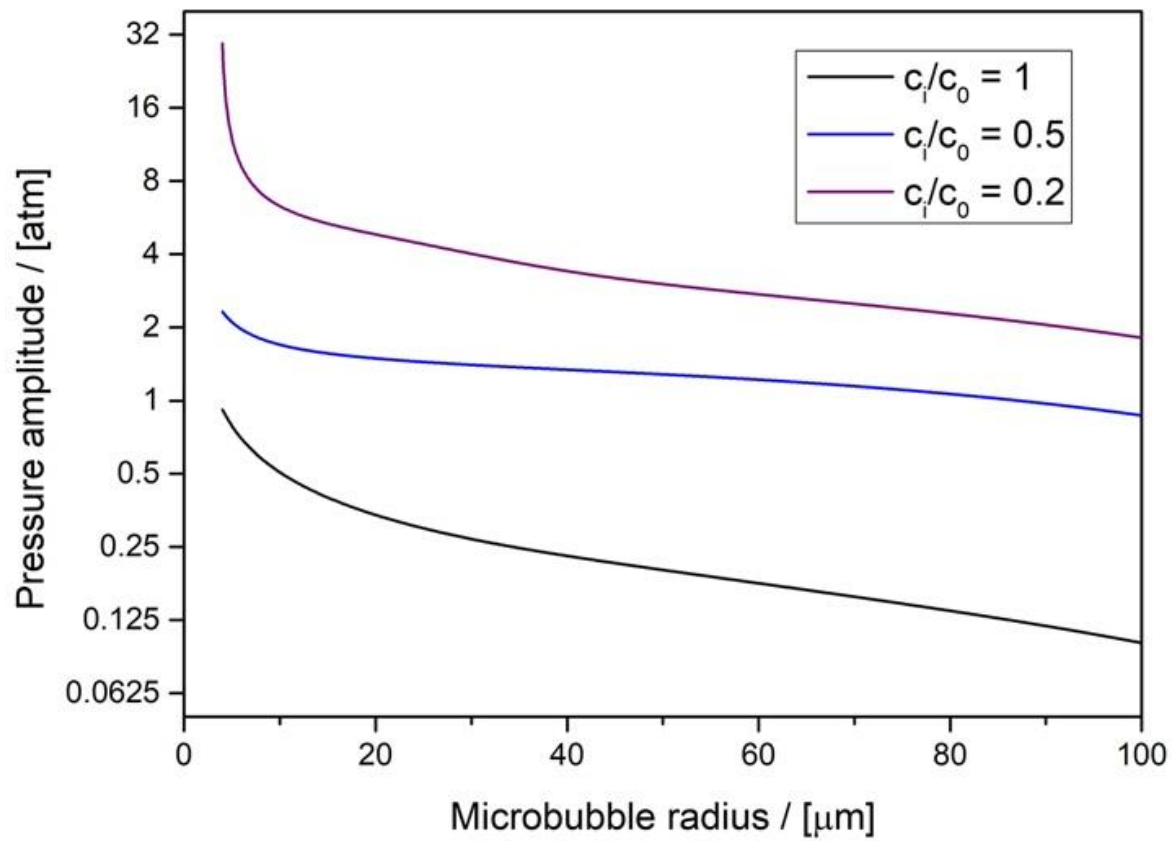

Fig. 11 The graphs of critical pressure $\tilde{p}_c$ against microbubble radius $R_0$

As is obtained the one-to-one relation between acoustic pressure $\tilde{p}$ and microbubble $R_0$ so we can get the size ranging of activated microbubbles under a given acoustic field. Referring to the work by Mellow [46, 47], the approximate analytic solution of acoustic field $\tilde{p}$ for the vibrating horn is available using the Green function method which formulas indicate that the distribution of $\tilde{p}$ relates to three variables, the horn radius $a$, the vibrating frequency $f$, and the vibrating amplitude $A_h$, but is only proportional to $A_h$ since the other two are fixed conditions in experiment. It is seen that the acoustic fields in Case A, B and C are identical, and become weaker in Case C and D sequentially for the vibrating amplitude reduced.

The diagrams in Fig. 12 illustrate the nondimensionalized analytic solution of acoustic field $\frac{\tilde{p}}{\rho_l c u_0}$ where the maximum value is located at the center of bottom wall and the semi-ellipsoidal iso-surfaces monotonically decrease down to the far field. It can be inferred from the background experimental pictures which represent the peak bulk of cavity that the region surrounded by the red iso-surface enables to provide sufficient driving force to develop the microbubbles evolving as cavitation nuclei. Thus, in term of the theoretical values of acoustic pressure $\tilde{p}$ inside the region, the span of qualified microbubbles, shown in Table 3, can be evaluated by Eqn. 38. According to the conclusions in Fig. 11, the ranges of $R_0$ display that the larger nuclei are appeared in the more degassed water (Case A, B and C), likewise situations embodied in weaker acoustic field (Case A and D).

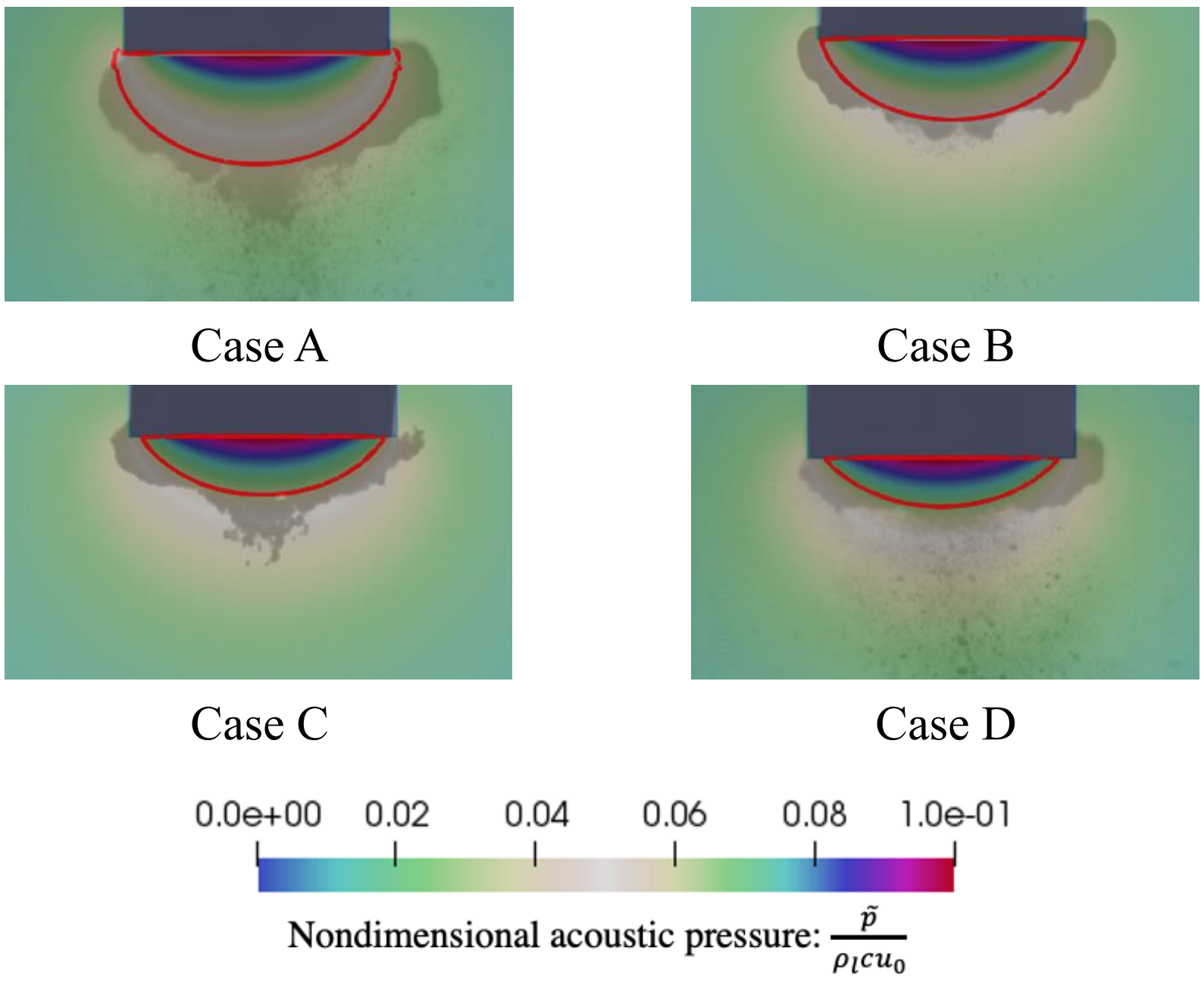

Fig. 12 The estimation of cavitation bubble production region

Since the value scope of microbubbles $R_0$ inside the effective region of acoustic field has been acquired, we suggest the multiples about 6~8 of $R_0$ to define the value range for model parameter $R_b$. And the range of maximum radius $R_M$ can exploit the numerical solution of Rayleigh-Plesset equation with the initial conditions of endpoint values of $\tilde{p}$ and $R_0$ that can estimate the parameter $R_m$. For clarity, all these specified values of five cases are listed in Table 3 as well.

Table 3 Estimation of model parameters

| | | Case A | Case B | Case C | Case D |
|---|---|---|---|---|---|
| Pressure amplitude in effective region $\tilde{p}$ $[bar]$ | | 6.18~15.8 | 7.72~15.8 | 9.27~15.8 | 6.12~9.63 |
| Activated microbubbles $R_0$ $[\mu m]$ | | 0.36~0.60 | 0.75~1.0 | 4.5~6.0 | 0.46~0.61 |
| Model parameters of $R_b$ $[\mu m]$ | Range | 2.2~4.8 | 4.5~8.0 | 27~48 | 2.8~4.9 |
| | Value | 3.5 | 5.2 | 35 | 4.6 |
| Maximum growth radius $R_M$ $[mm]$ | | 0.41~0.76 | 0.49~0.77 | 0.56~0.77 | 0.41~0.56 |
| Model parameters of $R_m$ $[mm]$ | | 0.58 | 0.62 | 0.66 | 0.48 |

4.2.2 Simulation setup

As is shown in Fig. 13, we employ a 2D axisymmetric computational domain consistently with the experiment with the dimensions of 40 mm height and 25 mm radius, in which the horn tip of 3 mm diameter is placed vertically from top 30 mm above the bottom. It is note that the near-wall grid is densified to ensure the y+ is lower than one.

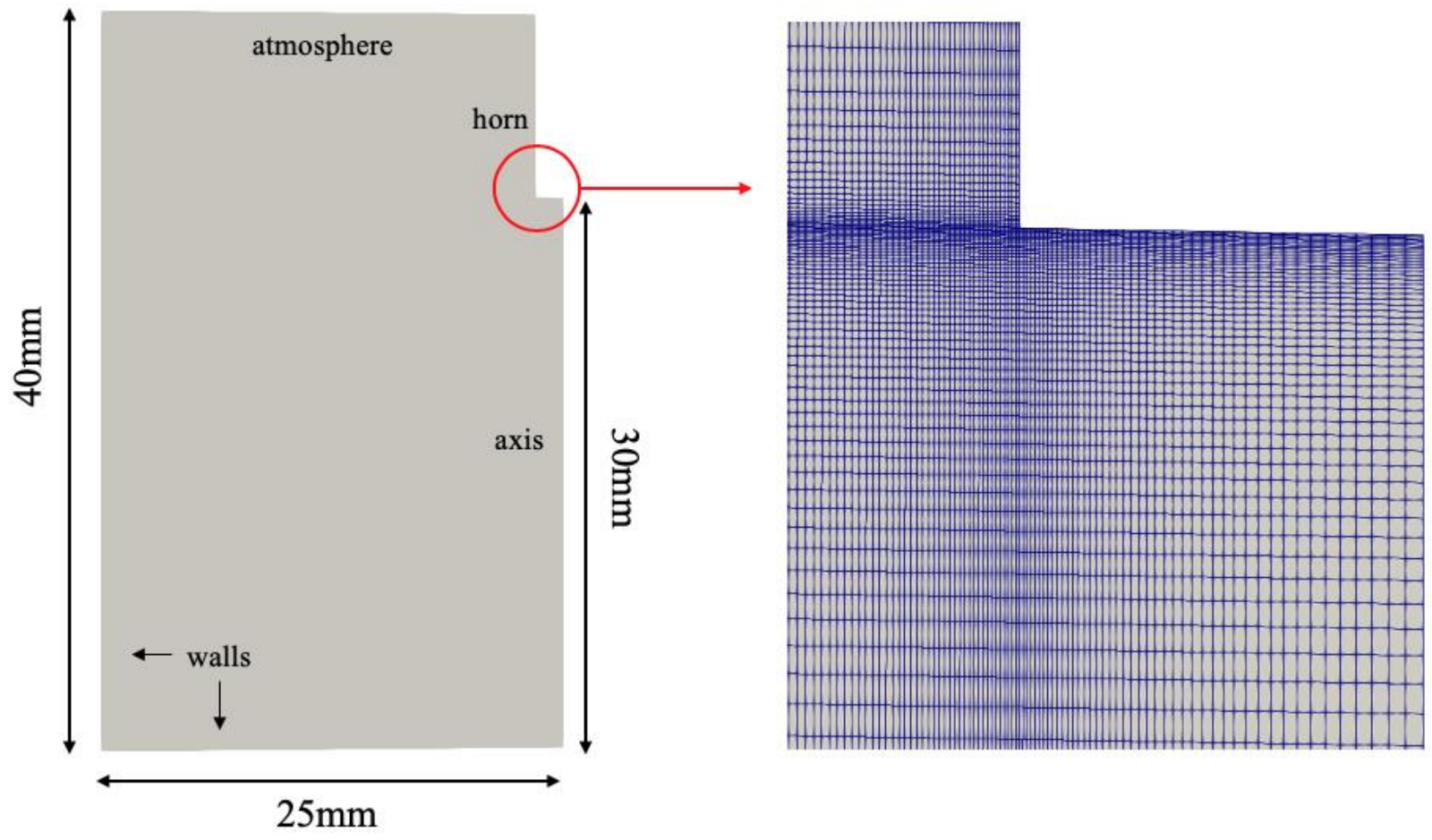

Fig. 13 Computational domain for ultrasonic horn

All the walls are used the no-slip velocity boundary condition and zero gradient for pressure. The top of atmosphere is defined as fixed pressure at 1 atm. The horn vibration in a sinusoidal manner at a frequency of 20 kHz, at various amplitudes, depending on the power. To capture the movement a dynamic mesh approach was used that the mesh must constantly be updated by laplacian smoothing and local remeshing. It was determined that the mesh, due to very small deformation of the domain, preserves an extremely low value of cell skew.

Three mesh densities were tested and it is found that it does not influence the outcome of the calculation of cavitation dynamics, but the model parameters have to change slightly for different resolutions. Consequently, the following results are calculated on a medium grid with 23550 cells.

### 4.2.3 Results

A series of simulations by the new cavitation model are compared with experiments including the volume evolution of the bubble cloud and the acoustic pressure probed by hydrophone. Considering that the physics of these cavitating flow do not differ significantly between Case A-D, we emphatically analyze the results of Case A whereas others are given more briefly in data charts.

A sequence of the spatial structures of cavity beneath the tip of the horn are shown in Fig. 14 graphically displayed as left simulation and right experiment. It is seen that the mushroom-like shape cloud is formed rapidly during the interval of 0 to $40\mu s$. Then, the generated bubbles keep the dynamic balance from $60\mu s$ to $100\mu s$ in which the maximum volume is achieved about $80\mu s$. After that the cavity contracts at the outer rim and the violent collapse happens at the end of cavitation period. A comparison between the measured and predicted cavity volume is shown in Fig. 15(a). It is evident that the simulation by NDCM accurately predicts the dynamics of the cavity volume which cycle (5010Hz) is about a quarter of the driving frequency (20kHz). Noting that the cavity frequency calculated by Schnerr-Sauer model can agree with the experiment, the produced vapor volume is insufficient though. The typical period inside pink dash line is magnified shown in Fig. 15(b), and attached the results obtained by Znidarcic's model additionally (dash line). We can see that the tendency is almost the same before $30\mu s$, however the shrink of cavity in growth period predicted by Znidarcic model mismatches with the measurements, moreover, the variation rate of cavity volume shifts much faster than the experiment during collapse process.

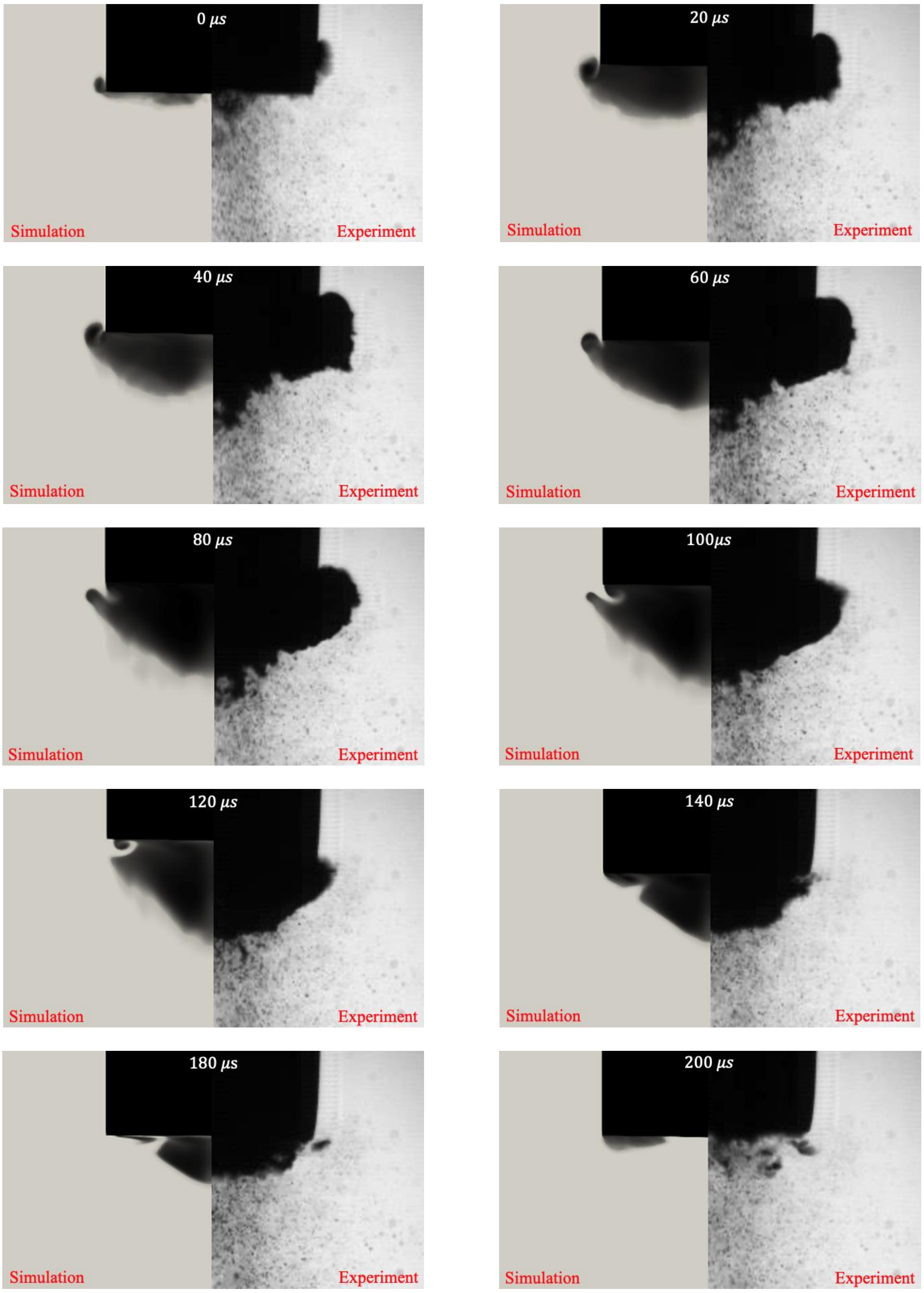

Fig. 14 Typical cycle of the oscillation of a large cavity between simulation and experiment at the driving frequency of 20kHz

Some discrepancies should be pointed out that a pinch of bubbles at the tip of cavity cannot capture perfectly possibly due to the 2D simulation method since the bubble cloud has non-axisymmetric 3D structure shown in experiment. Besides, the convection of corner bubbles which is likely induced by Bjerknes [48] force is unable to take into account here limited by the model capability.

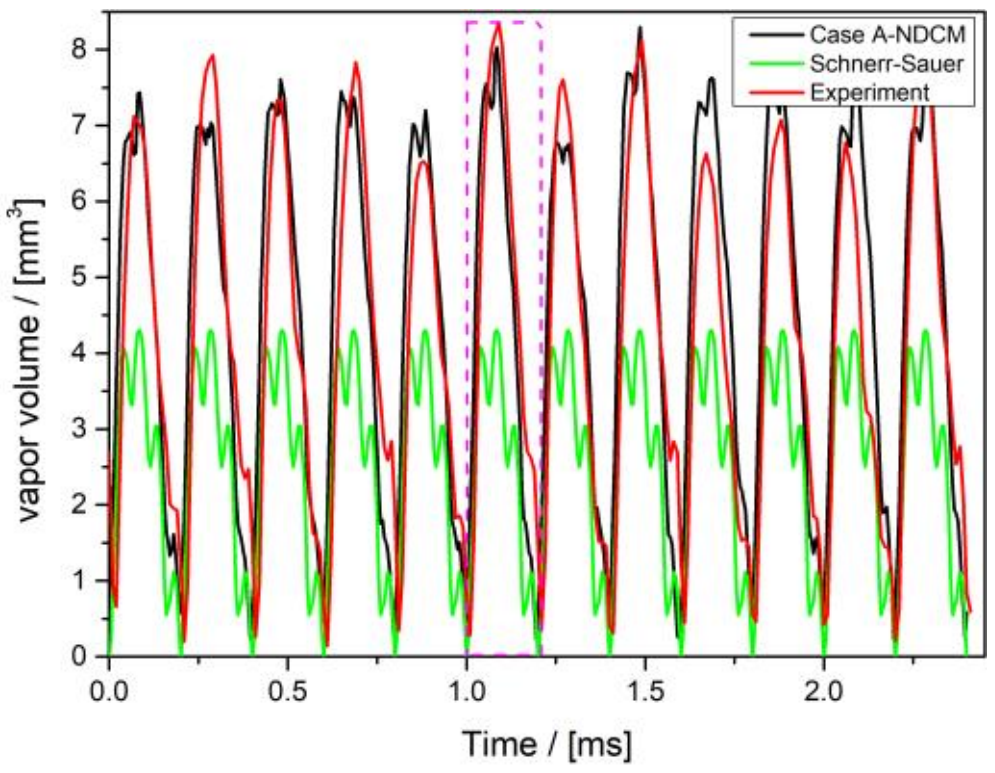

(a)

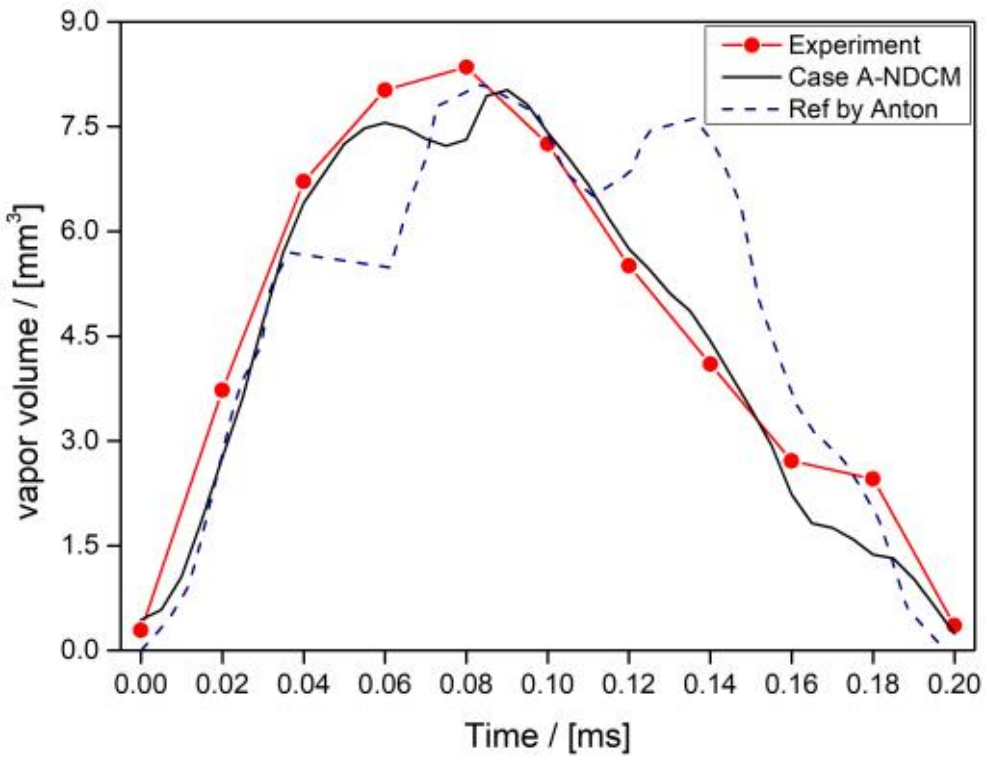

(b)

Fig. 15 Comparison between the predicted and measured attached cavity volumes for Case A

A comparison of measured and simulated pressure evolutions at a distance of 7 mm from the tip of the horn are shown in Fig. 16(a). It is indicated that high pressure pulse is emitted at the last stage of cavity collapse. The peak pressure amplitude seems to be slightly overpredicted which could be caused by the assumption of incompressibility for both phases, and furthermore missing the isolated bubble structures (see the right side in Fig. 14) where the cavitation model can only capture the main cavity bulks. Also note that the high frequency components are smoothed out because of the insufficient mesh resolution at the vicinity of probe and low order of temporal scheme under the RANS simulation scenarios. Nevertheless, the periodicity of pressure peaks is correctly predicted shown in Fig. 16(b) of power spectrum density (PSD). It is evident that the primary and second frequency in PSD are identical with the cavity and the driving horn respectively, which reflect main dynamic characteristics in the system.

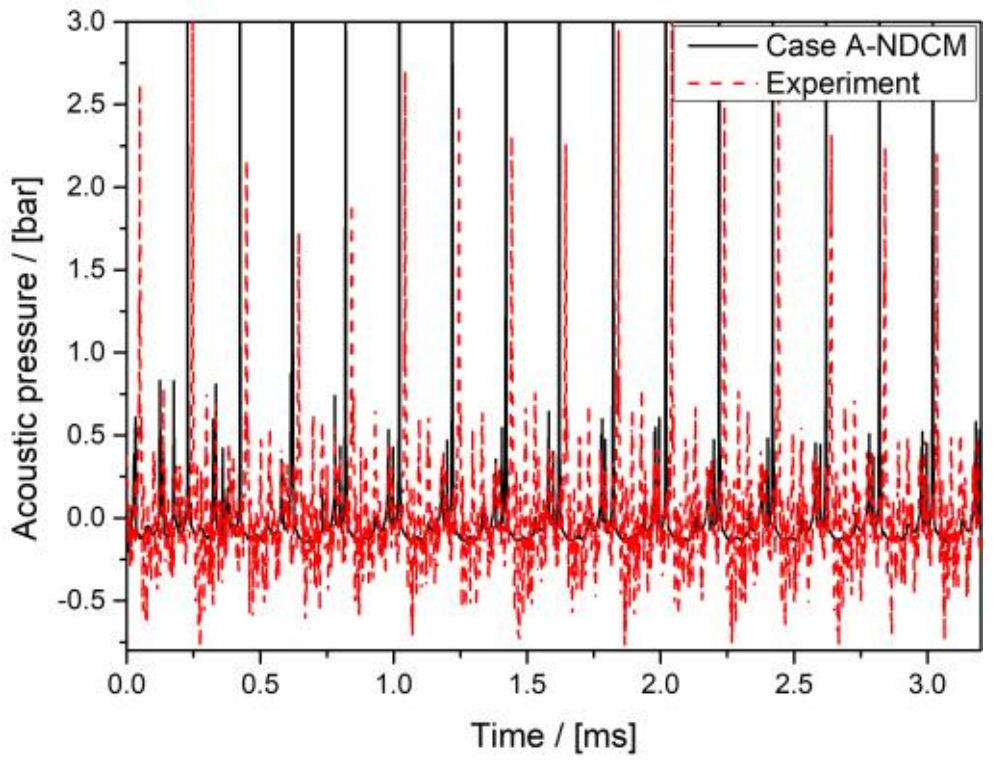

(a)

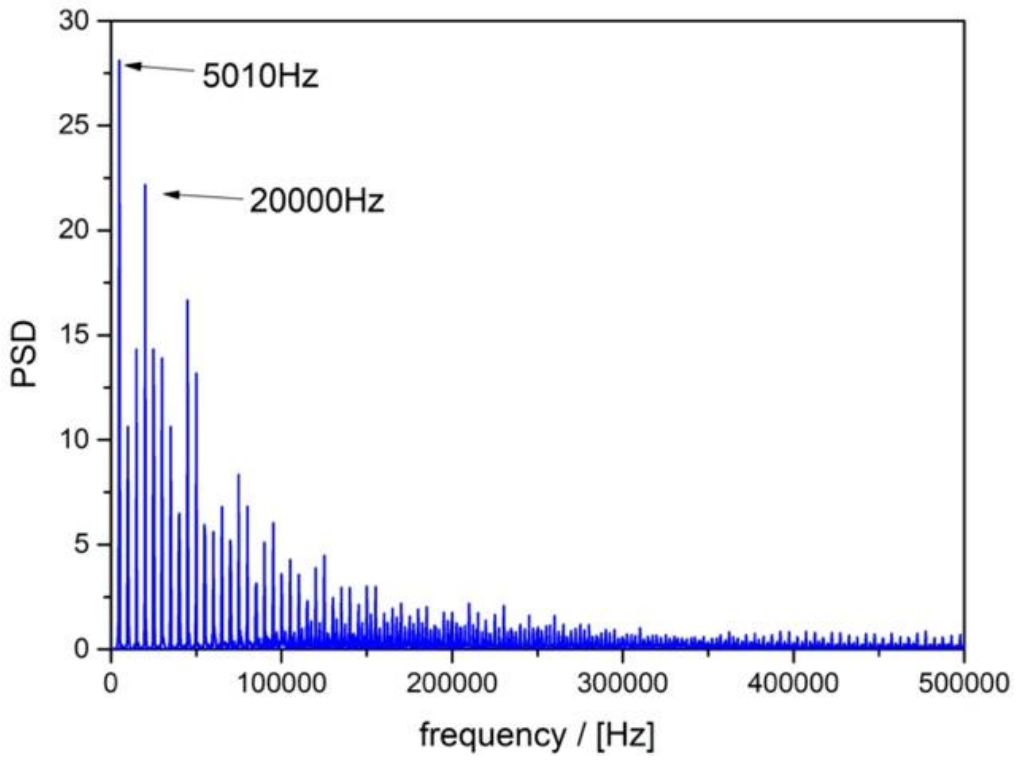

(b)

Fig. 16 Comparison between the predicted and measured acoustic pressure (a), PSD

analysis for the obtained acoustic pressure in Case A (b)

As unsuccessful application of Schnerr-Sauer model is well illustrated, we will discuss the rest of Cases B-D which model parameters are orderly defined in Table 3. The typical cavity volume evolution for degassed water of Case B and C are shown in Fig. 17(a) and Fig. 18(a) respectively, we can see that the peak volume are declining as the gas content decreased (about $7\sim8mm^3$ for Case A, $6\sim7mm^3$ for Case B, and $4\sim5mm^3$ for Case C), meanwhile accompany with increasing unsteadiness of the cavitation dynamics. This character can also be observed in acoustic pressure variations (Fig. 17 (b) and Fig. 18(b)) which amplitude wiggles violently by degassed extent. There exist complex bubble interactions inferred from Fig. 18(b) (Case C) that the pressure pulse radiates more frequently than the previous two where the calculated pressure reflects these frequency components qualitatively as well. However, the periodicity of the cavity does not damage which is slightly increased (5010Hz for Case A, 5079Hz for Case B, and 5259Hz for Case C) for more additional high frequency emerged. Comparing with Cases A, the peak volume is significantly reduced in Case D shown in Fig. 19(a) which is caused by the weak acoustic field by the low driving power. Differently the cyclic evolution period of cavity (6517Hz) is raised to about one-third of driving frequency.

It is seen that the NDCM is capable to predict the cavity dynamics accurately. Moreover, the proposed model parameters, $R_b$ and $R_m$, are physically based where good agreements can be obtained by setting reasonable values.

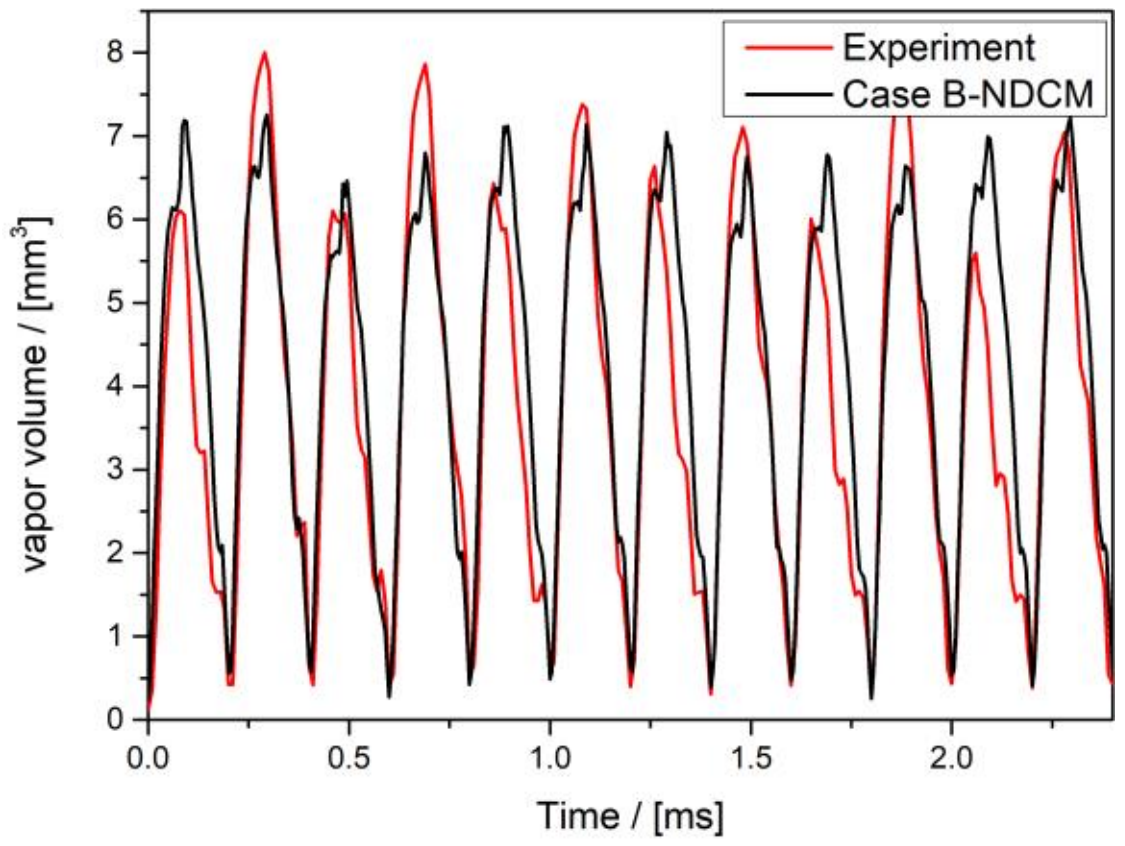

(a)

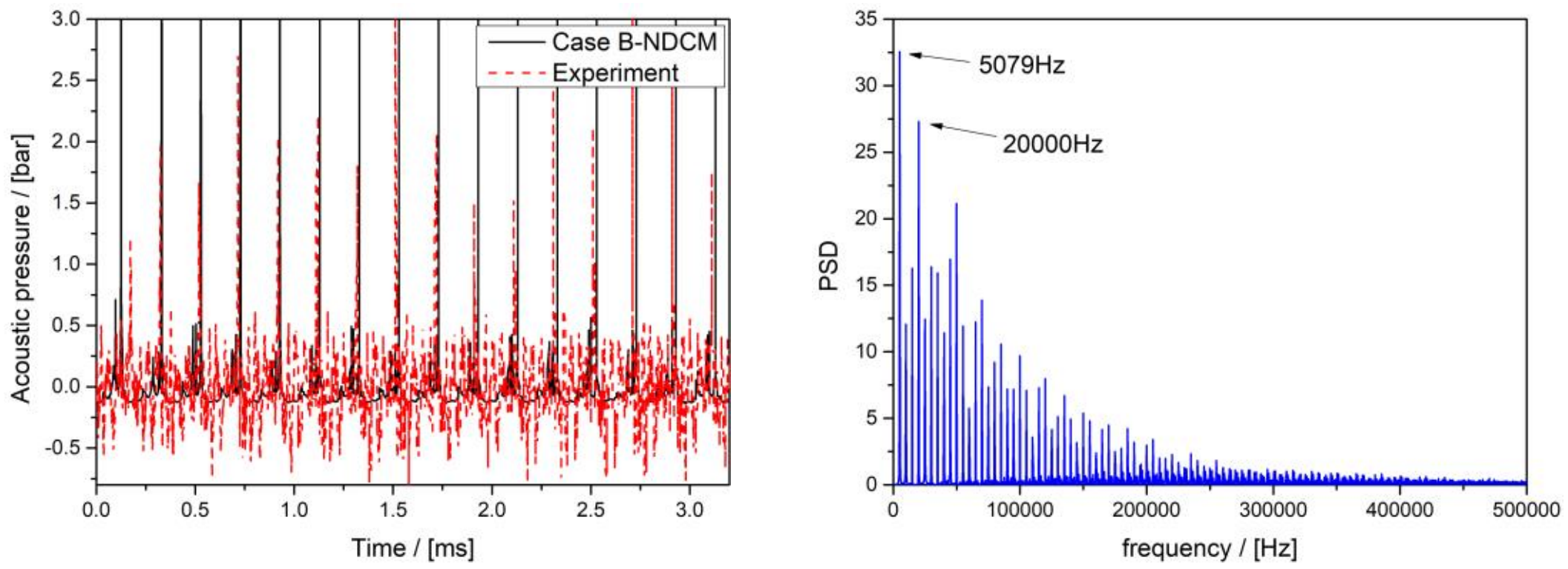

(b) (c)

Fig. 17 Comparison between the predicted results and measured data for Case B (70% max power, 50% saturated)

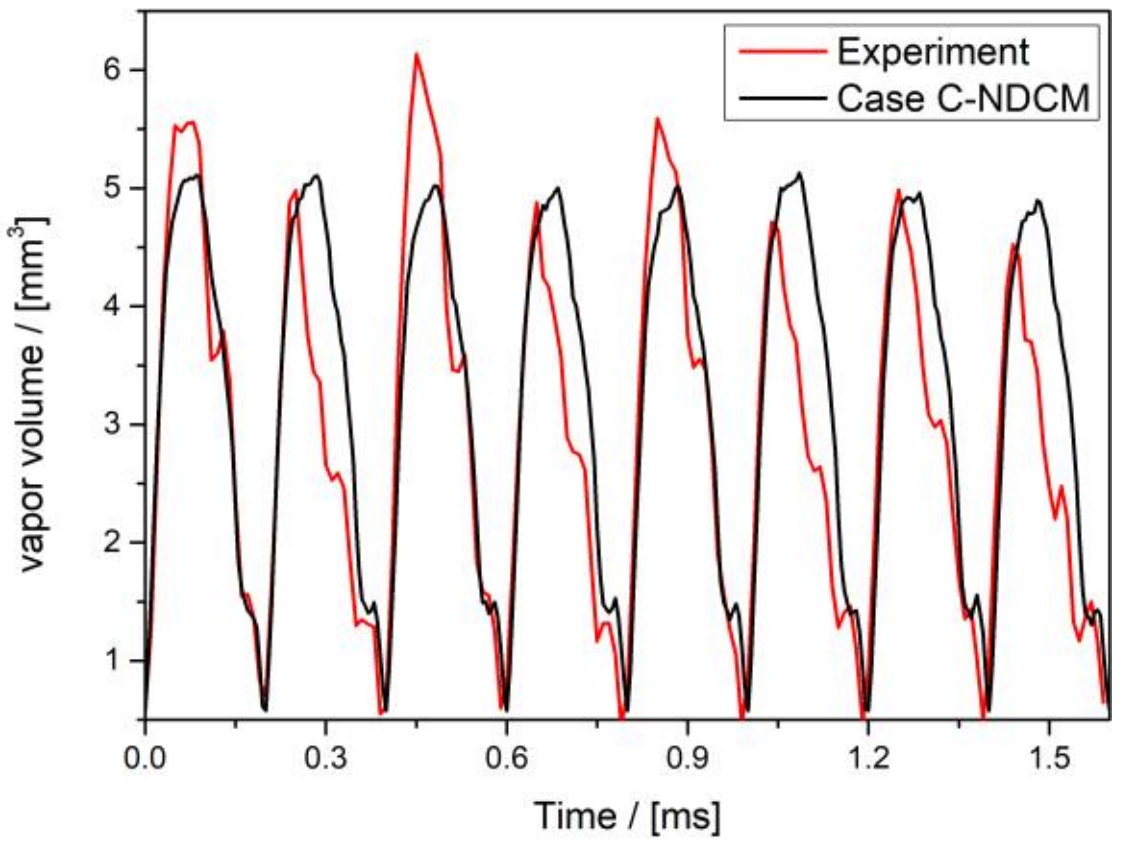

(a)

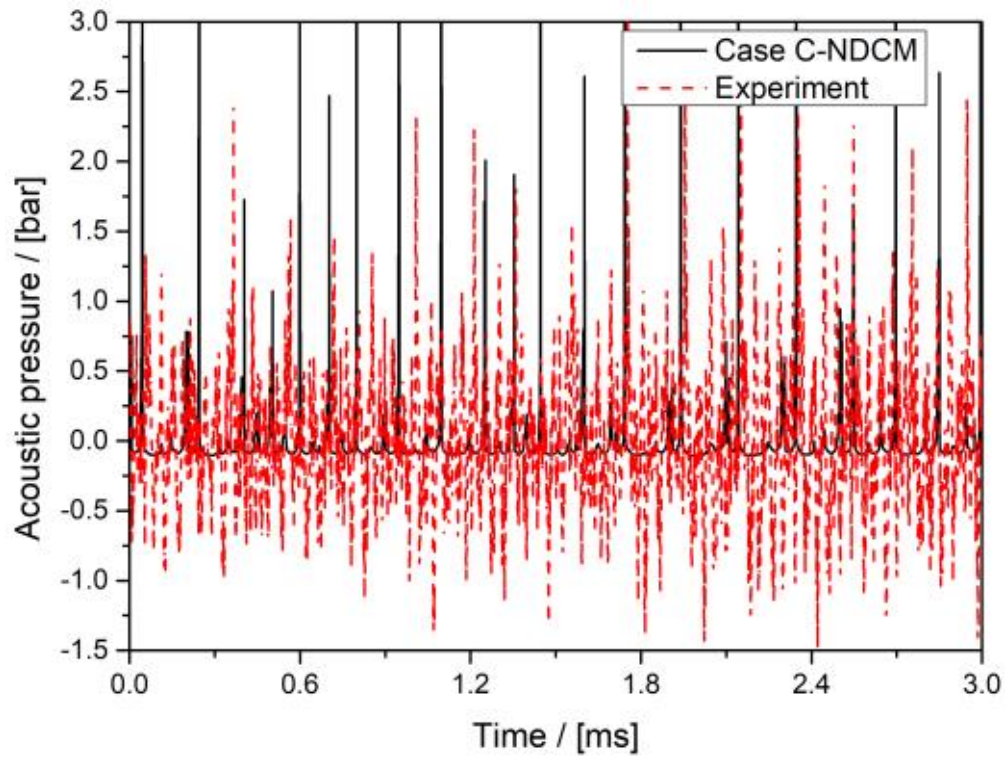

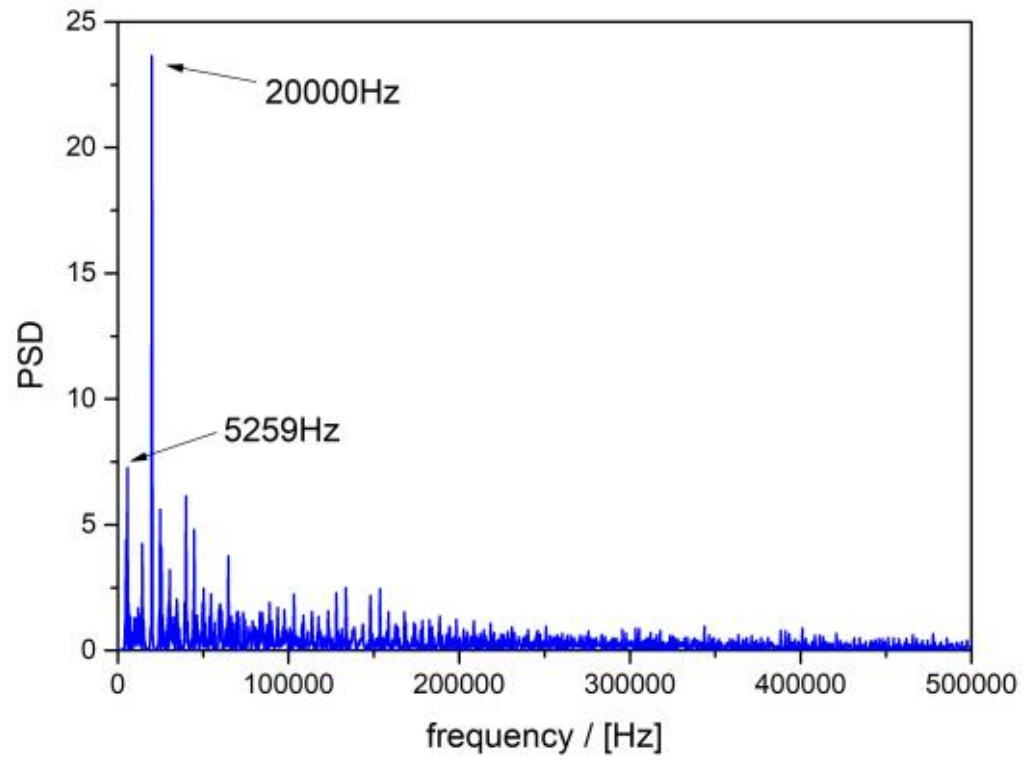

(b) (c)

Fig. 18 Comparison between the predicted results and measured data for Case C (70% max power, 20% saturated)

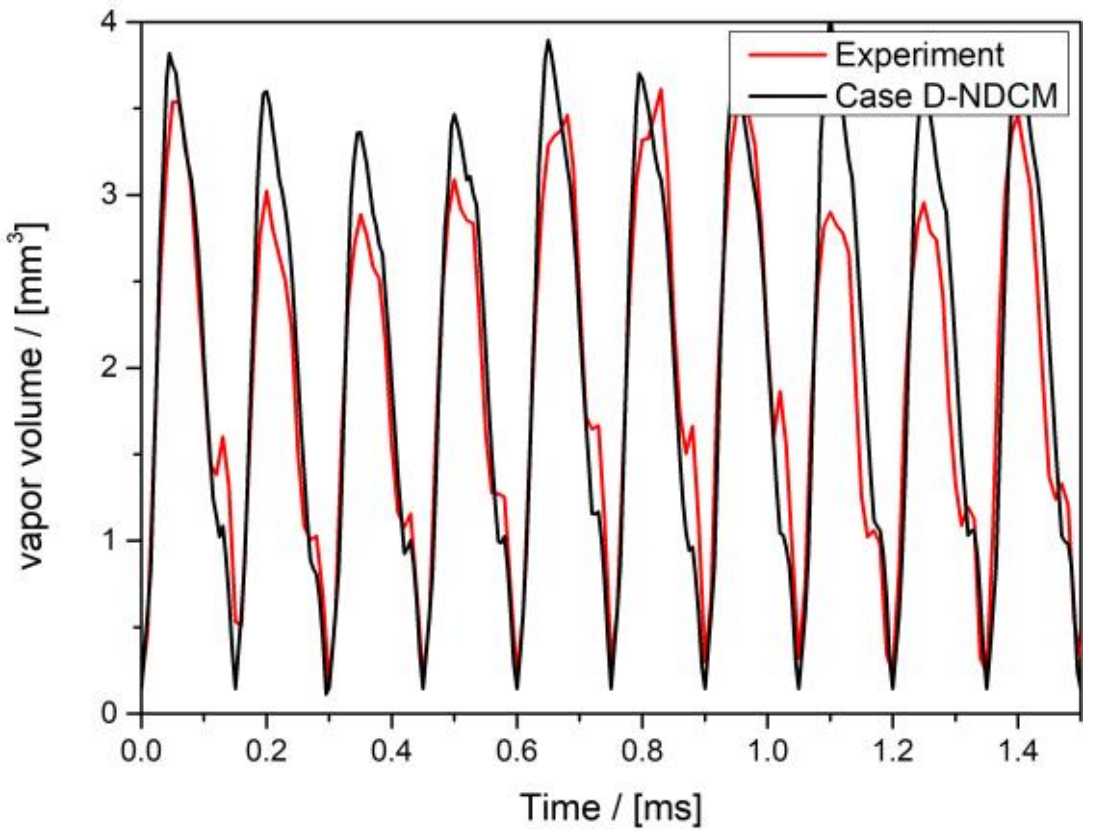

(a)

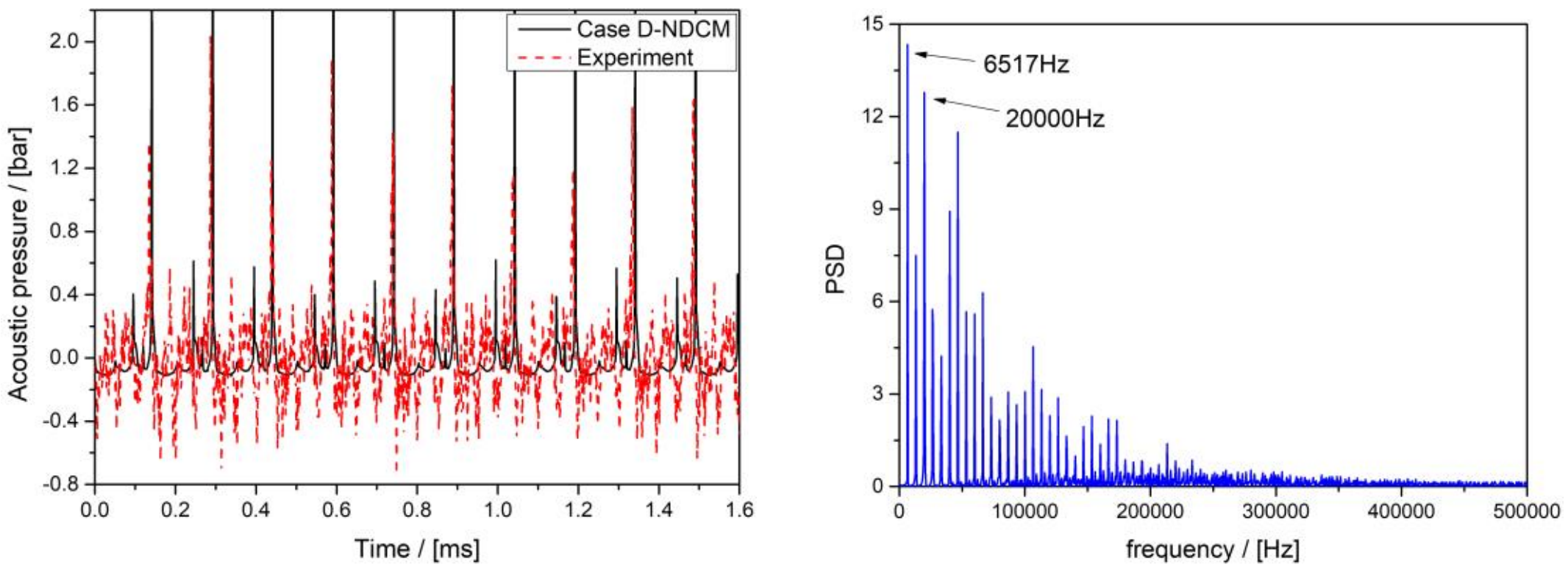

(b) (c)

Fig. 19 Comparison between the predicted results and measured data for Case D (30% max power, 100% saturated)

## 4.3 Slender body

In this section, we extend the application for NDCM on the hydrodynamic cavitation flow which contains the structures of dispersed bubbly cloud. The natural cavitation experiments [32] for axisymmetric bodies with blunt and conical heads are employed for elementary investigation, also to compare with the results by Merkle's model in the article.

The 2D axisymmetric computational domains are adopted shown in Fig. 20. The diameter of the slender body is 20 mm and the length 7.5d. The domain extension has 15d upstream and 20d downstream. The inlet velocity is 6.8m/s and fixed outlet pressure based on the cavitation number $\sigma$ with 0.5 and 0.3. The slender walls are specified as no-slip boundary and the outer ring is set as slip wall. The boundary layer grid is generated to ensure y+ less than 1.

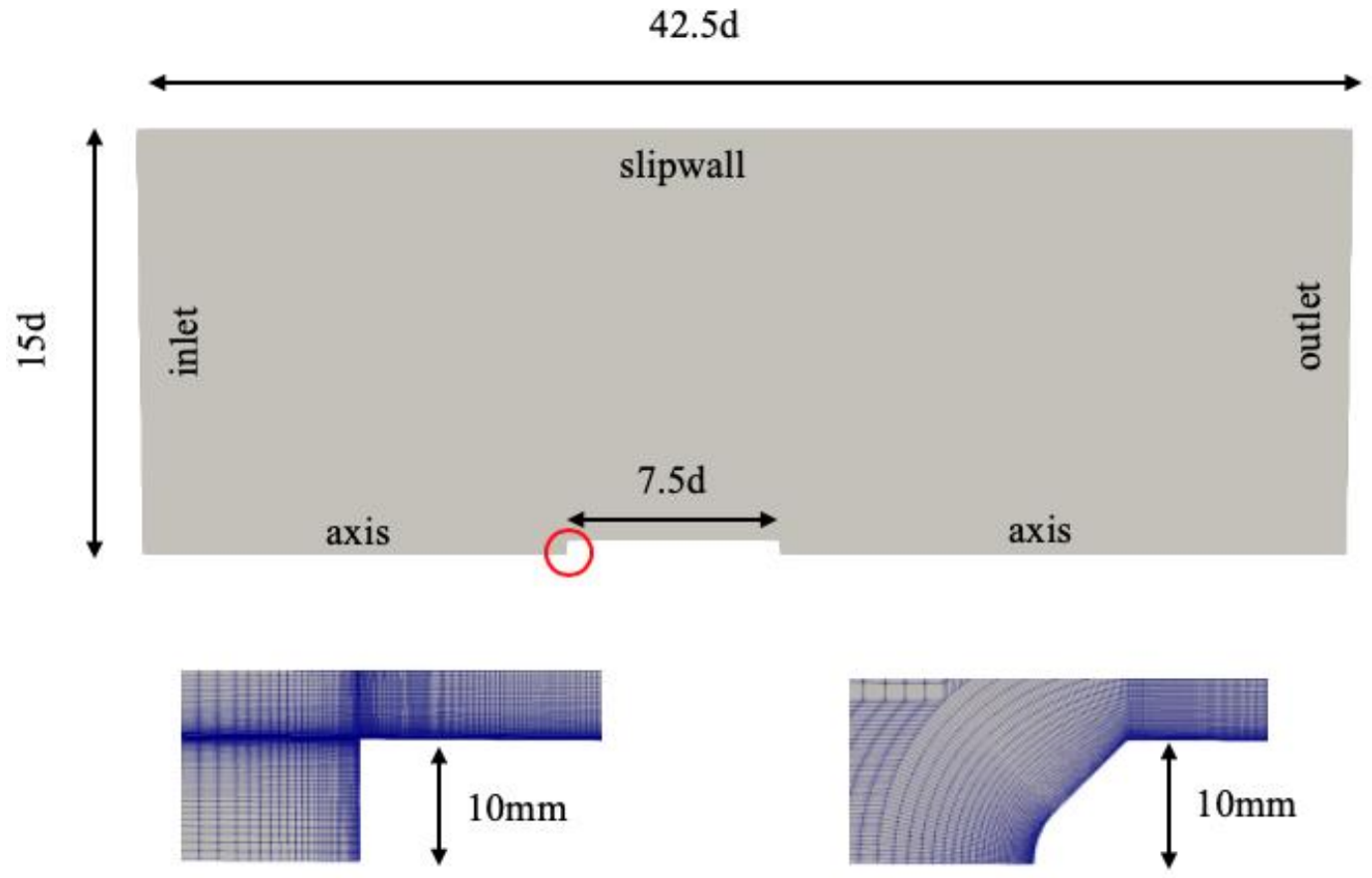

Fig. 20 Computational domain for slender bodies with conical and blunt heads

In order to determine the model parameters, it only needs to select the middle mesh with the quantity of 37700 cells and 45400 cells for conical and blunt heads respectively between three resolutions which the simulation results are close. For the lack of information to estimate the size of bubble nuclei, the approximate values of $R_b$ and $R_m$ for both heads listed in Table 4 are guessed from several trial calculations. We suppose that these values are reasonable because the cavitation region has higher negative pressure and wider area as dropping the cavitation number, which enables smaller nuclei to grow ($R_b$) and more residence time leads to bubbles expand larger ($R_m$).

Table 4 Model parameters of NDCM model

| | $\sigma = 0.5$ | $\sigma = 0.3$ |
|---|---|---|
| Blunt | $R_b = 32\ \mu m$<br>$R_m = 0.53\ mm$ | $R_b = 25\ \mu m$<br>$R_m = 0.62\ mm$ |
| Conical | $R_b = 28\mu m$<br>$R_m = 0.43mm$ | $R_b = 18\mu m$<br>$R_m = 0.52mm$ |

The pressure distributions for blunt head with cavitation number of 0.3 and 0.5 are shown in Fig. 21(a). We can see that the results by NDCM achieves good agreements especially for $\sigma = 0.5$ that the cavity length predicted by new model (solid line) is almost identical with the experiment but underestimated in Merkle's (dash line). The insufficient length of cavity predicted by Merkle's model is embodied notably in conical head shown in Fig. 21(b) whereas the NDCM presents a reasonable prediction matching with experiments.

However, some discrepancies still exist limited by 2D geometry that the recovery pressure at the cavity tail is a bit lower than experiment at $\sigma = 0.5$ and the cavity length is little overpredicted for blunt at $\sigma = 0.3$. It is more likely to use the 3D simulation to capture the asymmetric structures of bubble cloud which are shedding periodically that can obtain better results.

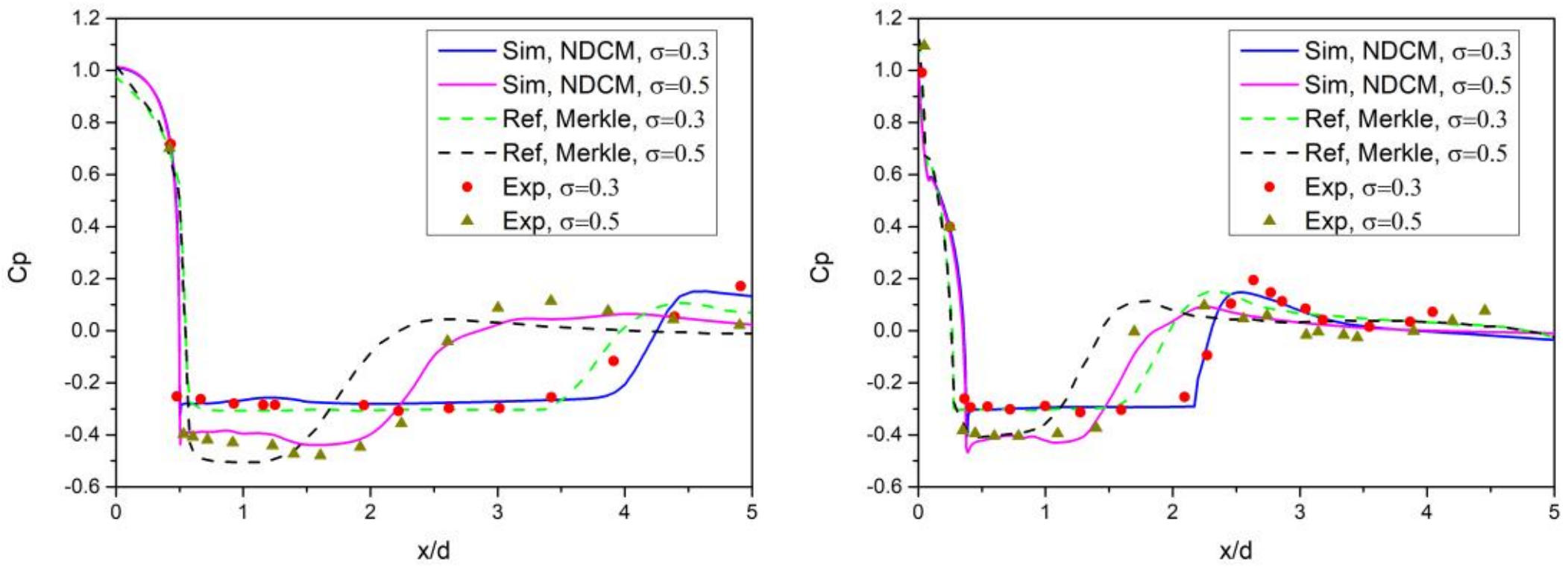

(a) blunt (b) conical

Fig. 21 Comparison of pressure distribution for blunt and conical head body

# 5. Conclusions

In this study, a novel nonlinear dynamic cavitation model (NDCM) is proposed against the bubble cluster structure through strictly mathematical derivation. Firstly, the four thorough assumptions for TFM-type models are explicated that the filtered bubbles are mapped from physical to computational space. Then, we introduced the integral average method to calculate the time derivative term $\frac{d}{dt}\left(\frac{4}{3}\pi nR^3\right)$ that the second derivative in Rayleigh-Plesset equation can be considered in the characteristic time during growth and collapse, namely $\tau_v$ and $\tau_c$, solved analytically. Consequently, two additional potential functions $\psi_v$ and $\psi_c$ emerge in model formula which represent the nonlinear effects in cavity dynamics. In addition, without any empirical coefficients, there are merely two parameters with definitude physical meaning in which $R_b$ and $R_m$ indicate the Blake critical radius and the average maximum growth radius, respectively.

In order to validate the performance of the new model, three simulation cases, from simple to complex, were employed including the collapse of numerical bubble cluster, periodic generation and collapse of real bubble cloud in ultrasonic horn experiment, and hydrodynamic cavitation of slender body.

For the first case, the results showed that the collapse time of NDCM and benchmark agreed well except the speed rate which may cause by different initialization method. The layer-by-layer collapse character and pressure shock at last stage were revealed correctly. On the contrary, the Schnerr-Sauer model with parameter $C_c = 1$ overpredicted the collapse time because of insufficient intensity of source term. More importantly, the potential function $\psi_c^S$ implied in Schnerr model gives a positive relation with bubble radius $R_b$ that contradicts with $\psi_c$ in NDCM. Although the collapse time can be remedied by employing large coefficient $C_c$, the model errors were also augmented that brought great numerical pressure wiggles and incorrect collapse processes. Most of spurious pressure can be suppressed by applying NDCM, but the remaining components should be further studied in future.

The NDCM was then applied to the real bubble cloud generated in acoustic field. The main purpose was to confirm the physical relevance of $R_b$ and $R_m$. Four experimental conditions were adopted that the theoretical value ranges of model parameters were determined based on the rectified diffusion theory. It was found that the variation of those well-matched simulation results was in accord with the laws of specified model parameters, and more sensitive to $R_b$ which should be given larger values in more degassed water or weaker acoustic field. A detailed comparison of Case A against the results from Znidarcic showed that the second derivative term in Rayleigh-Plesset equation considered in integral form rather than differential can provide better predictions.

Finally, the new model was extended to hydrodynamic cavitation of convective dominated flow that the slender bodies with two heads of conical and blunt were simulated under the cavitation number of 0.3 and 0.5. The good agreements of cavity length and pressure distribution further indicated that the NDCM is applicable for the cavitation cavity with dispersed bubble structures.

## Acknowledgement

This work is supported by National Key Project GJXM92579 and National Sci. & Tech. Major Project (2017-II-0004-0016).

## References

1. J.P. Franc, J.M. Michel, *Fundamentals of cavitation.* Fluid Mechanics & Its Applications, 2004. **76**(11): p. 1-46.
2. K. Maeda, T. Colonius, *Bubble cloud dynamics in an ultrasound field.* Journal of Fluid Mechanics, 2019. **862**: p. 1105-1134.
3. Y.C. Wang, C.E. Brennen, *Numerical computation of shock waves in a spherical cloud of cavitation bubbles.* Journal of Fluids Engineering, 1999. **121**(4): p. 872-880.
4. M. Brenner, S. Hilgenfeldt, D. Lohse, *Single-Bubble Sonoluminescence.* Reviews of Modern Physics, 2002. **74**(2).
5. W. Lauterborn, T. Kurz, *Physics of bubble oscillations.* Reports on Progress in Physics, 2010. **73**(10): p. 106501.
6. S. Fleckenstein, D. Bothe, *A Volume-of-Fluid-based numerical method for multi-component mass transfer with local volume changes.* Journal of Computational Physics, 2015. **301**: p. 35-58.
7. R.F. Kunz, D.A. Boger, D.R. Stinebring, T.S. Chyczewski, J.W. Lindau, H.J. Gibeling, S. Venkateswaran, T.R. Govindan, *A preconditioned Navier–Stokes method for two-phase flows with application to cavitation prediction.* Computers & Fluids, 2000. **29**(8): p. 849-875.
8. Y. Luo, J. Zhou, X. Yang, Z. Jiang, *A Pressure-Base One-Fluid Compressible Formulation for High Speed Two-Phase Flows With Heat and Mass Transfer.* Journal of Heat Transfer, 2018. **140**(8).
9. I. Senocak, S. Wei, *A Pressure-Based Method for Turbulent Cavitating Flow Computations.* Journal of Computational Physics, 2002. **176**(2): p. 363-383.
10. K. Maeda, T. Colonius, *Eulerian-Lagrangian method for simulation of cloud cavitation.* Journal of Computational Physics, 2018. **371**: p. 994-1017.
11. J.S. Ma, C.T. Hsiao, G.L. Chahine, *Euler–Lagrange Simulations of Bubble Cloud Dynamics Near a Wall.* Journal of Fluids Engineering, 2015. **137**(4).
12. A. Kubota, H. Kato, H. Yamaguchi, *A new modeling of cavitating flows: a numerical study of unsteady cavitation on a hydrofoil section.* Journal of Fluid Mechanics, 2006. **240**(240): p. 59-96.
13. D. Dauby, P. Queutey, A. Leroyer, M. Visonneau, *Computation of 2D cavitating flows and tip vortex flows with an unstructured RANSE solver.*, in *11èmes journées de l'hydrodynamique, Brest, France*. 2007.
14. R. Kunz, D. Boger, T. Chyczewski, D. Stinebring, H. Gibeling, T. Govindan. *Multi-phase CFD analysis of natural and ventilated cavitation about* *submerged bodies*. in *Proceedings of FEDSM '99*. 1999.
15. C.L. Merkle, J. Feng, P.E. Buelow. *Computational modeling of the dynamics of sheet cavitation*. in *3th International Symposium on Cavitation*. 1998.
16. M. Frobenius, R. Schilling, R. Bachert, B. Stoffel, G. Ludwig. *Three-dimensional unsteady cavitation effects on a single hydrofoil and in a radial pump : measurements and numerical simulation; P. 2: Numerical simulation*. in *International Symposium on Cavitation, Osaka, Japan*. 2003.
17. G.H. Schnerr, J. Sauer. *Physical and Numerical Modeling of Unsteady Cavitation Dynamics*. in *4th International Conference on Multiphase Flow, New Orleans, USA*. 2001.
18. A.K. Singhal, M.M. Athavale, H. Li, Y. Jiang, *Mathematical Basis and Validation of the Full Cavitation Model.* Journal of Fluids Engineering, 2002. **124**(3): p. 617.
19. P.J. Zwart, A.G. Gerber, T. Belamri. *A two-phase flow model for predicting cavitation dynamics*. in *Fifth international conference on multiphase flow, Yokohama, Japan*. 2004.
20. Y. Delannoy, J.L. Kueny. *Two phase flow approach in unsteady cavitation modelling*. in *Cavitation and Multiphase Flow Forum, ASME-FED*. 1990.
21. O. Coutier-Delgosha, R. Fortes-Patella, J.L. Reboud, *Evaluation of the Turbulence Model Influence on the Numerical Simulations of Unsteady Cavitation.* Journal of Fluids Engineering, 2003. **125**(1): p. 38-45.
22. M. Hofmann, H. Lohrberg, G. Ludwig, B. Stoffel, J. Reboud, R. Fortes-Patella. *Numerical and Experimental Investigations on the Self-Oscillating Behaviour of Cloud Cavitation and Part I: Visualisation*. in *3rd ASME/JSME Joint Fluids Engineering Conference, San Francisco, CA*. 1999.
23. C.G. Song, J.H. He. *Numerical simulation of cavitating flows by single-phase flow approach*. in *3rd International Symposium on Cavitation, Grenoble, France*. 1998.
24. M. Morgut, E. Nobile, I. Bilu, *Comparison of mass transfer models for the numerical prediction of sheet cavitation around a hydrofoil.* International Journal of Multiphase Flow, 2011. **37**(6): p. 620-626.
25. M. Morgut, E. Nobile, *Numerical Predictions of Cavitating Flow around Model Scale Propellers by CFD and Advanced Model Calibration.* International Journal of Rotating Machinery, 2012. **2012**: p. 618180.
26. T.D. Tran, B. Nennemann, T. Vu, F. Guibault, *Investigation of Cavitation Models for Steady and Unsteady Cavitating Flow Simulation.* International Journal of Fluid Machinery and Systems, 2015. **8**: p. 240-253.
27. A. Ducoin, B. Huang, Y.L. Young, *Numerical Modeling of Unsteady Cavitating Flows around a Stationary Hydrofoil.* International Journal of Rotating Machinery, 2012. **2012**: p. 1-17.
28. H. Jasak, *Error analysis and estimation for the finite volume method with applications to fluid flows.* Imperial College London, 1996. **m**(8): p. A385.
29. D. Ogloblina, S.J. Schmidt, N.A. Adams, P. Dančová, *Simulation and analysis of collapsing vapor-bubble clusters with special emphasis on potentially erosive* *impact loads at walls.* European Physical Journal Conferences, 2018. **180**.
30. A. Žnidarčič, R. Mettin, C. Cairós, M. Dular, *Attached cavitation at a small diameter ultrasonic horn tip.* Physics of Fluids, 2014. **26**(2): p. 023304.
31. A. Žnidarčič, R. Mettin, M. Dular, *Modeling cavitation in a rapidly changing pressure field – Application to a small ultrasonic horn.* Ultrasonics Sonochemistry, 2015. **22**: p. 482-492.
32. H.P. Wei, S. Fu, Q. Wu, B. Huang, G.Y. Wang, *Experimental and numerical research on cavitating flows around axisymmetric bodies.* Journal of Mechanical Science and Technology, 2014. **28**(11): p. 4527-4537.
33. S. Márquez Damián, *An Extended Mixture Model for the Simultaneous Treatment of Short and Long Scale Interfaces*. 2013.
34. H. Weller, *A New Approach to VOF-based Interface Capturing Methods for Incompressible and Compressible Flow*. 2008.
35. A. Cubero, N. Fueyo, *A Compact Momentum Interpolation Method for Unsteady Flows and Relaxation.* Numerical Heat Transfer, part B-Fundamentals, 2015. **56**: p. 507-529.
36. A. Cubero, A. Sanchez-Insa, N. Fueyo, *A consistent momentum interpolation method for steady and unsteady multiphase flows.* Computers & Chemical Engineering, 2014. **62**(mar.5): p. 96-107.
37. T. Du, J. Wang, Y. Wang, C. Huang, *A study of the collapse speed of bubble clusters.* International Journal of Multiphase Flow, 2020. **129**: p. 103322.
38. H. Ganesh, S.A. Makiharju, S.L. Ceccio, *Bubbly shock propagation as a mechanism for sheet-to-cloud transition of partial cavities.* Journal of Fluid Mechanics, 2016. **802**: p. 37-78.
39. V. Aeschlimann, S.P. Barre, H. Djeridi, *Velocity field analysis in an experimental cavitating mixing layer.* Physics of Fluids, 2011. **23**(5): p. 055105.
40. M. Florian, *Two-Equation Eddy-Viscosity Transport Turbulence Model for Engineering Applications.* AIAA Journal, 1994. **32**(8): p. 1598-1605.
41. C.M. Rhie, W.L. Chow, *Numerical study of the turbulent flow past an airfoil with trailing edge separation.* AIAA Journal, 1983. **21**(11): p. 1525-1532.
42. J. Klostermann, K. Schaake, R. Schwarze, *Numerical simulation of a single rising bubble by VOF with surface compression.* International Journal for Numerical Methods in Fluids, 2013. **71**(8): p. 960-982.
43. E. Ghahramani, M.H. Arabnejad, R.E. Bensow, *A comparative study between numerical methods in simulation of cavitating bubbles.* International Journal of Multiphase Flow, 2019. **111**: p. 339-359.
44. D. Rossinelli, B. Hejazialhosseini, P. Hadjidoukas, C. Bekas, A. Curioni, A. Bertsch, S. Futral, S.J. Schmidt, N.A. Adams, P. Koumoutsakos, *11 PFLOP/s simulations of cloud cavitation collapse*, in *Proceedings of the International Conference on High Performance Computing, Networking, Storage and Analysis*. 2013, Association for Computing Machinery: Denver, Colorado. p. Article 3.
45. L.A. Crum, *Acoustic cavitation series: Part five rectified diffusion.* Ultrasonics, 1984. **22**(5): p. 215-223.

46. T. Mellow, *On the sound field of a resilient disk in free space.* Journal of the Acoustical Society of America, 2008. **123**(4): p. 1880-1891.
47. T. Mellow, L. Kärkkäinen, *On the sound field of an oscillating disk in a finite open and closed circular baffle.* Journal of The Acoustical Society of America, 2005. **118**: p. 1311-1325.
48. F.G. Blake, *Bjerknes Forces in Stationary Sound Fields.* The Journal of the Acoustical Society of America, 1949. **21**(5): p. 551-551.